# The Innovative Distinctiveness of Prizewinners and their Networks

**Authors:** Chaolin Tian[1], Yurui Huang[1], Ching Jin[2], Yifang Ma[1*], Brian Uzzi[3*]

**Affiliations:**

[1]Department of Statistics and Data Science, Southern University of Science and Technology, Shenzhen, Guangdong, China

[2]Centre for Interdisciplinary Methodologies, University of Warwick, Coventry CV4 7AL, United Kingdom

[3]Northwestern Institute on Complex Systems (NICO) and Kellogg School of Management, Northwestern University, Evanston, Illinois, USA

**Emails: correspondence:** mayf@sustech.edu.cn; uzzi@kellogg.northwestern.edu

## Abstract

Science prizes purportedly reward innovation and explorations of new phenomena. Yet, in practice prizes may inadvertently divert resources from similarly impactful but less celebrated scholars. Despite this paradox, knowledge of how prizewinning relates to innovation is nascent even as prizes proliferate widely. Analyzing 2,460 worldwide prizes, we compared the innovativeness of over 23,000 prizewinners and matched non-prizewinners whose performance records were statistically equivalent up to the prize year. First, we find that prizewinners are more innovative. Their research is more likely to combine existing ideas in new ways, integrate a topic's historical and contemporary thinking, and incorporate interdisciplinary perspectives. Second, although prizewinners and matched non-prizewinners have statistically equivalent impact and productivity records up to the prize year, at about five years before the prize, prizewinners' papers become more innovative than their matched peers, a difference that widens each year, peaks during the prize year, and then persists for the remainder of their careers. Third, network embeddedness predicts unusual innovativeness. Compared to non-prizewinners, prizewinners' collaborations are shorter in duration, encompass wider exposure to unfamiliar topics, and involve coauthors whose networks minimally overlap with each other. The implications of the findings for the efficacy of reward systems and innovation in science are discussed.

## Significance Statement

This study compares the innovativeness of thousands of science prizes worldwide, their prizewinners, and a matched sample of similarly impactful non-prizewinners. We find that scientific prizewinners are more likely to produce papers that combine existing ideas in new ways, join foundational and current research on a topic with regards to emerging problems, and are more interdisciplinary than their matched non-prizewinning peers. Prizewinners' exceptional innovativeness appears linked to their unique network embeddedness. Relative to non-prizewinners, prizewinner's collaborations have shorter durations, more exposure to new topics, and lower overlap among collaborators.

## Introduction

Science prizewinners disproportionately influence scientific thought, resources, problems, and policies ([1-15](#)). They are celebrated in the media, at conventions, in journals and announcements, and in hallway conversations and nomination letters highlighting a scholar's work and reputation ([1](#), [2](#), [5](#), [16-22](#)). Their work is associated with an unexpected surge in the number of scholars who pivot their work to the prizewinning topic ([5](#)) and their ideas spread across the walls of scientific disciplines more broadly than equally cited, non-prizewinning ideas ([14](#)). Prizewinners' protégés are prone to become prizewinners themselves, creating lineages of influence that can span generations ([12](#), [23-27](#)).

While prizewinners are heralded by scientists, governments, and the public, research has raised concerns about the influence of prizewinners on science. The 41st chair problem posits that there are typically more equally good contenders for a prize than there are prizes ([1](#)), which potentially leads to equally strong ideas not receiving equal attention. Relatedly, as prizes have proliferated over time, they have become more concentrated among fewer prizewinners, raising concerns that a diminishing fraction of innovative ideas are gaining attention ([14](#)), which can potentially limit the breadth of ideas in science ([14](#)) and can underrepresent different demographic groups ([2](#), [15](#), [26](#), [28-30](#)). Consistent with these concerns, researchers and policy analysts argue that more research is needed on understanding leadership in science and the conditions that favor scientific innovation ([4](#), [31-33](#)).

Here, we address questions about whether prizewinning scientists differ in their innovativeness relative to equivalently productive and impactful non-prizewinning contenders. We use science prizes as a measure of a shared community belief that the prizewinner's work -- though equivalent in citation impact and productivity to their peers -- is exceptionally innovative ([2](#)). Our data includes information on over 2,460 prizes that were conferred on 7,353 worldwide prizewinners. In conjunction with the prize dataset, we created two more original datasets. The second dataset is made up of up to five matched non-prizewinners who are from the same field and career stage, and who have citation and productivity records that are statistically indistinguishable from the prizewinners' records up to the year of the prize, for a total sample of 23,562 prizewinners and matched non-prizewinners. The third dataset measures the embeddedness of the collaboration networks of prizewinners and matched non-prizewinners ([34-36](#)).

**Data**

We assembled data on 2,460 international scientific prizes and their 7,353 recipients (1900-2018) and merged the information from official prize websites and Wikipedia. Prizewinner data includes awards and award dates, research topics studied, and publications that significantly expanded prior prize datasets (5, 14). The data used to compute the innovation measures was curated from OpenAlex (37) via its citation network records. OpenAlex employed an advanced author name disambiguation algorithm to identify prizewinners by using their names, publication records, affiliations, citation patterns, and external identifiers like ORCID (support information (SI) Sec. 1, Table S1, and S2 report descriptive information on the combined prize/prizewinner dataset).

**Innovativeness Measures**

To quantify a range of accepted operationalizations of innovativeness used in science, we employed three popular, replicable, and validated measures (38). Our first measure is "novelty". Novelty measures the degree to which a scholar's papers combine prior knowledge in conventional or novel ways (39). Novel papers combine past knowledge in ways that have not been or rarely have been combined before. Conversely, conventional papers combine knowledge in familiar ways (36, 40-42). Quantitatively, the more a paper combines the work listed in its bibliography in ways that are less than expected by chance, the more it combines ideas in new or rarely seen-before ways. Conversely, the more a paper combines the work referenced in its bibliography in ways that are greater than expected by chance, the more it combines ideas in familiar, seen-before ways (see SI Sec. 2.1 for details).

Our second measure is "convergence" (32, 43). Convergence measures the degree to which a paper integrates foundational and recent ideas on a topic (7, 32, 33, 43-46). By linking foundational and recent ideas on a topic, high convergence papers join a topic's original thinking with new thinking and findings, inventively adapting historical ideas to modern problems or emerging applications (32). Quantitatively, to capture the integration of historically foundational ideas and the latest thinking on a topic, convergence is computed as the mean and coefficient of variation (CV) of the publication years of a paper's references with respect to the focal paper's publication year. For each focal paper, we calculate the age differences between the publication year of the focal paper and all the papers it references and then compute the mean and CV of ages. When the age difference between the focal paper's publication year and its references' publication years has a low mean age but a high CV, the focal paper connects historical and present-day insights on a topic (see SI Sec. 2.2 for details).

Our third measure of innovativeness is "interdisciplinarity" (36, 47-51). Interdisciplinarity measures the degree to which a scholar's work incorporates disparate subject categories; here we use the level-1 concept in OpenAlex as the subject categories. Quantitatively, we compute a focal paper's interdisciplinarity as: $\Delta = \sum_{ij(i \neq j)} d_{ij} p_i p_j$, where $p_i$ and $p_j$ are the proportion of the focal paper's citing papers in subject categories $i$ and $j$. $d_{ij}$ is the cosine dissimilarity score between $i$ and $j$, which is calculated through the co-citation matrix for all papers. The value $\Delta$ ranges from 0 to 1, and higher values indicate that the paper is more interdisciplinary (see SI Sec. 2.3 for details).

To ensure that the three measures load on the same underlying construct, we conducted a principal component analysis (PCA). Novelty, convergence, and interdisciplinarity loaded in distinct directions within a broader, multidimensional space. The PCA indicated that these measures capture related but different aspects of innovation ([38]), which is consistent with the low bivariate correlations between novelty and convergence (0.018), novelty and interdisciplinarity (0.035), and convergence and interdisciplinarity (0.037) (see SI Sec. 2.4, Table S3, and Figure S4 for details).

**Matching Procedure**

We utilized coarsened exact matching (CEM) ([52]) and dynamic optimal matching (DOM) ([53]) methods to construct comparable groups of prizewinners (PWs) and matched non-prizewinners (NPWs) up to the year each PW received their first award. CEM matches PWs and NPWs based on fixed categorical variables. DOM matches PWs and NPWs based on their time-varying attributes on a year-by-year basis. For example, if total citations are spread over five publications a year, DOM looks for matches where the distribution of citations over papers in a single year is equivalent between PWs and NPWs to account for surges in productivity or high-impact publications. The variables selected for matching include these six characteristics: (a) discipline, (b) first publication year (research age), (c) total number of publications before the prizewinning year, (d) total number of citations before the prizewinning year, (e) yearly number of publications, and (f) citations before the prizewinning year.

For each prizewinner, according to the CEM procedure, we identify their first publication year, first prizewinning year, research discipline, total number of publications, and the total number of citations to create initial super groups of matched PWs and NPWs ($\leq 200$ for each PW) where the first year of publication can vary by $\pm 2$ years and total publications and citations can vary by $\pm 30\%$ relative to the target PW before their prizewinning year. Within each initial super group, we used DOM to further select the NPWs whose dynamic fluctuations in yearly publications and citations were statistically indistinguishable from the PWs. We define the distance measure ($d_{i,j}^p$) for PW $i$ and the NPW $j$ based on the yearly number of publications and yearly number of citations ($d_{i,j}^c$) $t_0$ years before the prizewinning year, as follows,

$$d_{i,j}^p = \frac{1}{t_0 + 1} \sum_{t=t^*-t_0}^{t^*} \frac{|Y_i(t) - Y_j(t)|}{Y_i(t)}$$

where $Y_i(t)$ presents the number of publications of PW $i$ at year $t$, $t^*$ represents the prizewinning year of PW $i$, and we traced back the matching for $t_0$ years before the prizewinning year. The same definition is used for distance $d^c_{i,j}$ for the number of citations (see SI Sec. 3.1.1 and Figure 1(A) for the procedure details). To ensure the closeness and balance between the PWs and NPWs for the entire system, for each PW, we selected up to five matched NPWs from the super-group pool who have the smallest distance within a distance threshold (SI Sec.3.1.2, Figures S7, and S8 use different thresholds and confirm the robustness of our results). NPWs not meeting the threshold were omitted from the analysis. Figure 1 (B-C) shows that the raw data on the number of papers and the number of citations for the PWs and NPWs with 95% CIs have no statistically significant differences before the prizewinning year.

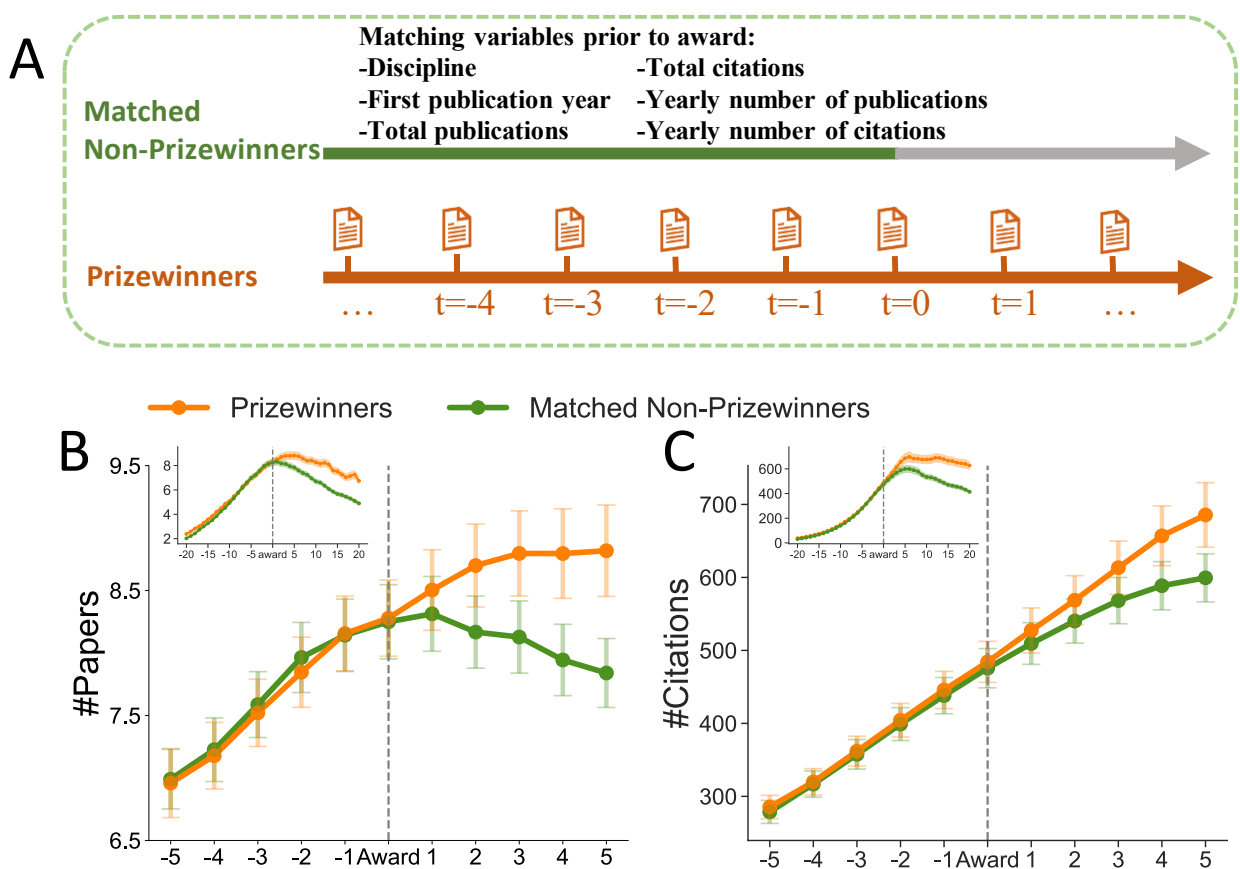

**Figure 1. Matching Process and Validation**. (A) Procedures for matching prizewinners with up to five non-prizewinners who have statistically indistinguishable impact and productivity records with the prizewinning scientists before the prizewinning year using CEM and DOM techniques. (B) and (C) confirm matching for productivity (# papers) and productivity (# citations) with 95% CIs. Insets show data for 20 years before and after the prize year. t=0 is the prizewinning year.

Numerous tests confirmed the robustness of the matching. First, SI Sec. 3.1.2, Figures S7 and S8 indicate that the results are robust to different thresholds. Second, "Mahalanobis" distance tests ([54-56](#)) in the DOM procedure demonstrated that results are robust to different distance methods (see SI Sec. 3.2.2, Sec. 4.5.1, Figure S10, and Table S8; SI Sec. 3.2.3, Table S4, $T$-tests all $p$-values > 0.5 and all SMD tests < 0.1). Third, the above results are robust to adding team size (see SI Sec.

3.2.1 and Figure S9) and dynamically matching the three innovativeness variables (see SI Sec. 3.2.4 and Figure S11).

**Innovation Gap**

Figure 2 (A-C) reports the raw data values of papers for PWs and NPWs for each innovativeness measure. The x-axes show career time. The prizewinning year is designated as $t^*$ on the x-axis, and the y-axes show the percentage of innovative papers written by PWs, NPWs, and a random sample of scholars (for each prizewinner, we randomly select 5 non-prizewinning scholars in the same discipline) on three separate plots for (A) novelty, (B) convergence, and (C) interdisciplinarity with 95% CIs. Figure insets show the same relationships when the innovation variables are included in the dynamic matching to ensure that the inclusion or exclusion of these variables do not alter the results.

The plots reveal several notable observations about the innovativeness of different classes of scholars over a career. First, for all three outcome variables, PWs' and NPWs' curves run parallel to each other. This pattern indicates that while PWs and NPWs differ in their levels of innovativeness, their trends in innovativeness mirror one another -- broadly moving at the same time and in the same direction. For example, when PWs' innovativeness rises or falls, NPWs broadly parallel those changes except for the period just prior to the prize year when the innovativeness of PWs begins to exceed that of NPWs. By contrast, relative to PWs and NPWs, a random sample of scholars follows a different pattern of innovativeness and their level of innovativeness is always significantly below that of PWs and NPWs.

Second, while PWs' and NPWs' growth and decline trends in innovativeness within any particular form of innovation mirror each other, the trends differ across innovation measures. Novelty (Figure 2(A)) continually increases over a career. By contrast, convergence and interdisciplinarity (Figures 2(B) and 2(C)) appear relatively flat during a scholar's early to mid-career stage and then relentlessly decline over the remainder of a career.

Third, just prior to the prize year, the innovativeness of PWs and NPWs begin to diverge. About four-to-five years before the prize, PWs' innovativeness begins to exceed that of NPWs and gradually and continually widens up to the prize year at which point the difference in innovativeness peaks. After the prize year, the innovation gap between PWs and NPWs stabilizes at roughly the peak level, and then persists until the end of their careers. This result suggests that on average scholar's innovativeness precedes their prizewinning and that prizewinners' relatively high innovativeness persists over the remainder of their careers. Relatedly, it is noteworthy that when a PWs' greater level of innovativeness emerges before the prize year, it occurs when PWs' and NPWs' productivity and impact are statistically indistinguishable (Figure 1). This result indicates that PWs initially diverge in innovativeness not by increasing their productivity or citation rates relative to NPWs but by publishing more innovative work at levels of productivity and impact comparable to matched NPWs. Thus, PWs innovativeness increases on a paper-by-paper basis prior to the prize. Figure 2 insets show the same analysis when the innovation variables are added to the matching process, which guarantees equivalence in innovation before the prize. The insets show that PWs diverge from NPWs in terms of the innovativeness of their papers, a gap that widens up to the prize year and roughly persists at the peak level thereafter.

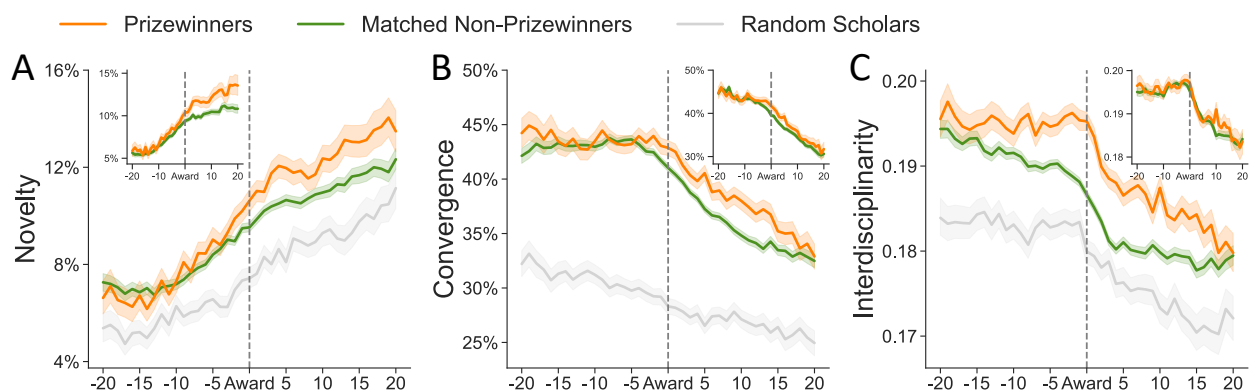

**Figure 2. Observed Innovativeness of Prizewinners, Matched Non-Prizewinners, and a Random Sample Group.** The y-axis shows the (A) percentage of novel papers, (B) percentage of convergence papers, and (C) mean interdisciplinarity values for prizewinners (PWs), non-prizewinners (NPWs) and a random sample of authors with 95% confidence intervals. The prizewinning year is designated on the x-axis as zero. The values of innovativeness for PWs and NPWs mirror one another over their careers and show no statistical differences in innovativeness about four-to-five years before the prize when PWs' and NPWs' innovativeness diverges and grows, peaks at the prize year, and then persists on all three measures of innovativeness over the remainder of their careers. Insets show the same relationships when the innovation variables are dynamically matched in the DOM procedure.

To more precisely quantify the estimated differences between PWs and NPWs, we separately regressed each innovation variable on an indicator variable for prizewinning (1=Yes), an interaction term for *prizewinner×post*, and controls variables including team size, matched group, prize, publication year, and author position. This regression resembles a difference-in-difference regression ([38](#)) and is used here to estimate statistical rather than causal relationships and to confirm the parallel trends requirement of PWs and NPWs before prizewinning (shown in SI Sec. 4.2) ([57](#)). We avoid making causal inferences because the proxy "treatment" variable, namely prizewinning, is not random but partially predictable through media coverage, informal conversations, or bibliographic data ([27](#), [58](#)).

In Figure 3, Panel A reports the regression results comparing PWs' and NPWs' innovativeness. Consistent with the raw data findings, the regression variable *Prizewinner* confirms that PWs display higher levels of research innovation than NPWs in terms of novelty, convergence, and interdisciplinarity ($p<0.05$, $p<0.01$, and $p<0.001$, respectively). The interaction term *Prizewinner×post* indicates that PWs relative to NPWs display significantly wider gap in their research novelty and convergence after the prize year ($p<0.001$ for both variables). By contrast, the insignificant interaction of *Prizewinner×post* for interdisciplinarity indicates that PWs produce more interdisciplinary research than NPWs, but the innovation gap for interdisciplinary before and after the prize does not differ in magnitude. Figure 3, Panel B, presents margin plots based on the regression models. We observe that PWs and NPWs display overlapping levels of innovativeness during their formative career years, with PWs showing a gradually widening and significant research innovation gap over NPWs that first emerges roughly in the five years before the prize year, peaks at the prize year, and then persists at that level after the prize year.

The SI presents several robustness checks. First, Table S5 reports robustness check for separate disciplines; Figure S13 reports robustness check when the sample is split into high and non-high prestigious prizes; and Table S6 reports robustness checks when the data is split into the periods before and after 1990 or just the last 30 years of data (see SI Sec.2.4 and Figure S5(A-I)). Second, we conduct goodness of fit replication regression analysis using nonlinear random-forest models (see SI Sec.4.5.5), Figure S15 reports actual vs predicted value plots, and staggered DID models are presented in SI Sec. 4.5.4 and Table S11). Additional specifications that control for PWs who win multiple prizes (see SI Sec. 4.5.2 and Table S9) and include or omit review papers confirmed the main results (see SI Sec. 4.5.3, and Table S10 for details).

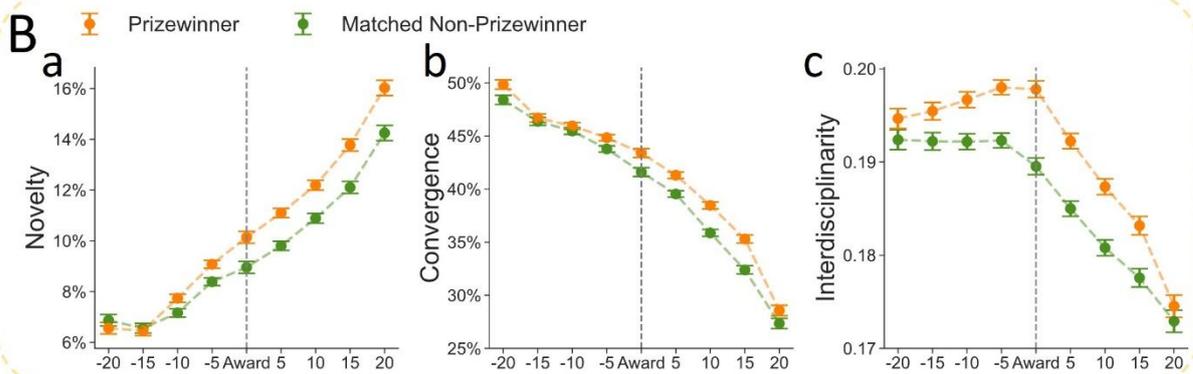

| | Model 1 | Model 2 | Model 3 |
|---|---|---|---|
| | **Novelty** | **Convergence** | **Interdisciplinarity** |
| **Prizewinner** | 0.003* (0.002) | 0.009** (0.003) | 0.004*** (0.001) |
| **Prizewinner×post** | 0.012*** (0.002) | 0.012*** (0.003) | 0.001 (0.001) |
| **Novelty** | -- | <-0.001 (0.001) | 0.014*** (0.000) |
| **Convergence** | <-0.001 (0.000) | -- | 0.002*** (0.000) |
| **Interdisciplinarity** | 0.217*** (0.005) | 0.067*** (0.008) | -- |
| **Prize** | Y | Y | Y |
| **Team size** | Y | Y | Y |
| **Matched group** | Y | Y | Y |
| **Author position** | Y | Y | Y |
| **Publication year** | Y | Y | Y |
| **Constant** | 0.055*** (0.001) | 0.386*** (0.002) | 0.184*** (0.000) |
| **Observations** | 2,838,611 | 2,838,611 | 2,838,611 |
| **R-squared** | 0.037 | 0.065 | 0.140 |

Robust standard errors in parentheses, *** p<0.001, ** p<0.01, * p<0.05.

**Figure 3. Regression Estimates of Prizewinners' and Matched Non-Prizewinners' Innovativeness. (A)** Each model controls for the fixed effects of PWs and NPWs matched groups, team size (six categories, 1, 2, 3, 4, 5, >5), prize, publication year (relative to the prizewinning year), and author position (see SI Sec. 4.3 for the model specifications). **(B)** Figures a-c represent the model-estimated innovativeness by career years (95% CIs). The estimated growth and decline patterns of PWs and NPWs mirror one another over their careers except for interdisciplinary. At the boundary of roughly five years before the prize, PWs gradually grow more innovative than NPWs, a gap that peaks at the prize year and continues to for the remainder of their careers (see SI Sec. 4.4 for the model specifications).

## Embeddedness and Innovation

Prizewinners' collaboration networks and innovativeness may be interrelated. The embeddedness of a researcher's network has been found to be positively associated with innovation through access to information, trust-building, and a decrease in process losses. (36, 59-62). Networks have also been found to be negatively related to innovation by creating echo chambers of like-mindedness (35, 59, 61, 63, 64). To examine embeddedness' potential role in the innovativeness of PWs and NPWs, we operationalized the embeddedness of PW's, NPW's, and the random sample of scholars' co-authorship networks using three embeddedness measures: tie duration, tie overlap, and topic similarity (35, 59, 65-68). Tie duration quantifies the average length of a co-authorship relationship. If a scholar collaborates on a paper that was published in 1995 with two coauthors (e.g., A and B), and the scholar's first collaboration record with A was in 1990, and with B was in 1991, then the duration of the ties associated with the 1995 paper is 4.5 years (9 total years divided by the two coauthors) (see SI Sec.5.1, and Figure S16(A)). Tie overlap uses the Jaccard index to measure the size of the intersection of coauthors divided by the size of their union ($J$). If, before publishing a joint paper with coauthors A and B, coauthor A had six total collaborators and B had eight, of which three overlapped, then $J$=3÷11≈0.273 (see SI Sec.5.1, and Figure S16(B)). Topic similarity quantifies the similarity between a scholar's pre-research topics and the topic(s) of the focal paper. For instance, if a scholar has, prior to the focal paper, published papers covering 12 distinct topics and the focal paper encompasses four topics of which two appeared in prizewinner's prior research topics, the topic similarity value for the focal paper would be 0.17 (=2/12) (see SI Sec.5.1, and Figure S16(C)). We used the fine-scale concepts classifications in OpenAlex to identify distinct topics.

Figure 4 presents the raw data for our embeddedness variables for PWs, NPWs, and the random sample of authors. The x-axes and y-axes represent career time and levels of tie duration, tie overlap, and topic similarity (95% CI shown) respectively. Over time, PWs', NPWs', and random authors have levels of embeddedness that move in parallel. PWs on average have lower levels of embeddedness than NPWs, and NPWs have lower levels of embeddedness than random authors on all measures. Further, topic similarity and tie overlap are positively correlated with each other and negatively correlated with tie duration, suggesting the main difference in the embeddedness of PWs and NPWs networks is more a matter of magnitude than structural form.

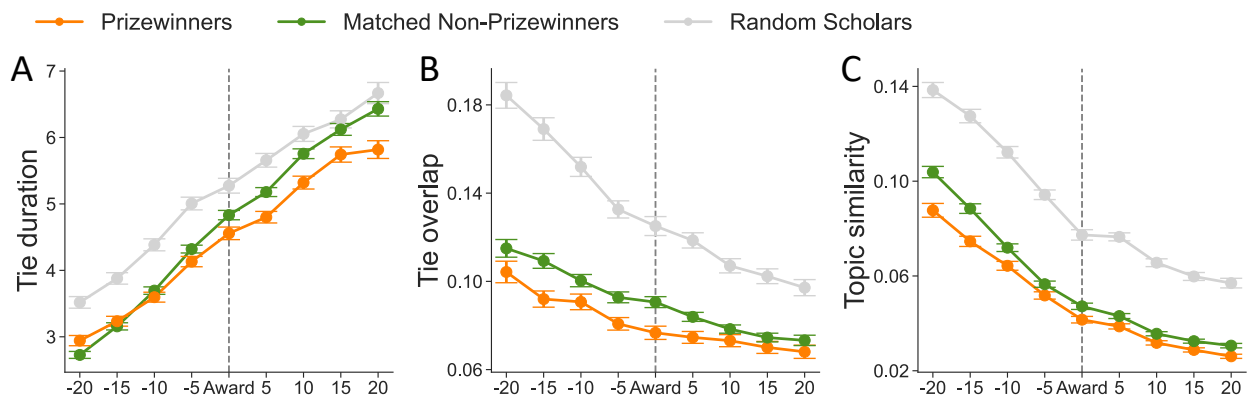

**Figure 4. Network Embeddedness Dynamics.** The dynamics of tie duration (A), tie overlap (B), and topic similarity (C) over the scholars' careers. Dots denote the averages, and error bars represent the 95% confidence intervals. Orange lines are for PWs, green lines are for NPWs, and gray lines are for random scholars.

Table 1 regresses our three innovativeness variables on the network embeddedness of PWs and NPWs. The regressions control for whether the focal scholar is a PW (1=Yes), pre- and post-prize periods (1=Yes), and fixed effects for prize, team size, publication year, matched group, and author position (see SI Sec. 5.2 for model specifications). The regressions reveal several links between embeddedness, prizewinning, and innovation. First, for both PWs and NPWs network embeddedness predicts innovativeness. Except for the one case out of nine coefficients, tie duration, tie overlap, and tie similarity are all highly significant and inversely associated with a scholar's novelty, convergence, and interdisciplinary. Second, the significant relationships between embeddedness and innovation when all embeddedness variables are simultaneously controlled in the model indicate that their relationships with innovation are additive. For example, the lower the tie duration, tie overlap, and topic similarity, the greater a PW or NPW's innovation. Third, after accounting for collaboration network patterns, the coefficients for *Prizewinner* are still positive and significant. These results suggest that prizewinning is not fully accounted for by a scholar's network of collaboration. To test whether these differences in relative innovativeness are due to PWs coauthors being more productive and cited than NPWs, we explored the productivity and citation impact of the coauthors of PWs and NPWs in our data. Figure S17 shows that over careers, the coauthors of PWs and NPWs have no statistical differences in their productivity and impact. Consistent with our inferences that the association between networks and innovation is not related to differences in the productivity and impact of PWs and NPWs coauthors.

**Table 1. Prizewinner's Network Embeddedness and Innovation.** Each model controls for the fixed effects of PWs and NPWs matched groups, team size (six categories, 1, 2, 3, 4, 5, >5), prize, publication year (relative to the prizewinning year), and author position (see SI Sec. 5.2 for model specifications). First, tie duration, tie overlap, and topic similarity are significantly and inversely related to novelty, convergence, and interdisciplinary, except for one case, convergence and topic similarity. After accounting for a scholar's embeddedness, prizewinners are still more innovative than matched non-prizewinners, suggesting that individual characteristics of scholars and their embeddedness in collaboration networks are associated with the level of innovativeness.

|  | Model 1 | Model 2 | Model 3 |
|---|---|---|---|
|  | **Novelty** | **Convergence** | **Interdisciplinarity** |
| **Prizewinner** | 0.005* (0.002) | 0.008* (0.004) | 0.006*** (0.001) |
| **Prizewinner × post** | 0.013*** (0.003) | 0.012** (0.004) | <-0.001 (0.001) |
| **Tie duration** | -0.001*** (0.000) | -0.015*** (0.000) | -0.001*** (0.000) |
| **Tie overlap** | -0.018*** (0.003) | -0.014** (0.005) | -0.026*** (0.001) |
| **Topic similarity** | -0.034** (0.011) | -0.016 (0.020) | -0.250*** (0.005) |
| **Novelty** | -- | -0.001(0.002) | 0.013*** (0.000) |
| **Convergence** | <-0.001 (0.001) | -- | 0.002*** (0.000) |
| **Interdisciplinarity** | 0.239***(0.006) | 0.075*** (0.010) | -- |
| **Prize** | Y | Y | Y |
| **Team size** | Y | Y | Y |
| **Matched group** | Y | Y | Y |
| **Author position** | Y | Y | Y |
| **Publication year** | Y | Y | Y |
| **Constant** | 0.071*** (0.001) | 0.489*** (0.002) | 0.194*** (0.000) |
| **Observations** | 1,750,607 | 1,750,607 | 1,750,607 |
| **R-squared** | 0.039 | 0.069 | 0.167 |

Robust standard errors in parentheses, *** p<0.001, ** p<0.01, * p<0.05.

## Discussion

Prizewinners disproportionately influence science. With prizes proliferating worldwide and the effectiveness of science's traditional rewards being debated, a key question is how prizewinners' innovativeness compares with that of less celebrated scientists. We quantify and compare the innovativeness of 23,562 PWs and matched NPWs who had statistically indistinguishable impact and productivity records before the prize year. Our results showed that PWs are significantly more innovative than NPWs. Prizewinners are more likely to combine existing knowledge in novel ways, connect foundational and cutting-edge ideas on a topic, and draw on interdisciplinary perspectives. Also, we found that a distinguishing predictor of the prizewinners' greater innovativeness is their embeddedness in collaborative relationships. In contrast to matched non-prizewinners, prizewinners have shorter-term collaborations, engage more often with research areas that are new to them, and have less overlap with their collaborators' collaborators. Dynamically, the level of innovativeness of PWs and NPWs begins to diverge significantly between the two groups at about five years before the prize, then widens consistently until the prize year when the gap peaks and thereafter stabilizes at roughly the peak level over the remainder of their careers.

The implications of our analysis for reward systems are associated with the Matthew Effect, a concept that holds that in practice prizewinners receive more rewards and recognition – fame, funds, or collaborators – than their peers regardless of the actual quality of their current work ([1](#)). Aiming to extend prior research on scientific rewards systems, our work examined both pre-prize and post-prize Matthew Effects. In contrast to prior research, our analysis uses a large sample of diverse prizes worldwide, rather than a single prize. Further, we created a five-to-one sample of dynamically matched non-prizewinners who had the same career age, working in the same discipline, and had statistically indistinguishable records of productivity and impact with each prizewinner.

The implications of our investigations relate to thinking on Matthew Effects, prizewinners, and prizes in two primary ways. First, our pre-prize results demonstrate that although prizewinners and matched non-prizewinners have no statistical differences in their productivity or impact up to the prize year, in years four-to-five before the prize, prizewinners grow increasingly more innovative than matched non-prizewinners. They publish significantly more work that combines existing research in new ways, integrates historical and contemporary ideas on a topic, and is more interdisciplinary. This result suggests that among equally outstanding researchers, greater innovation in research is significantly associated with prizewinning.

Second, to examine post-prize Matthew Effects, we tested whether prizewinners experience cumulative advantages relative to non-prizewinners after receiving their prize. To examine this relationship, we split our prize data into high and low prestige prizes under the assumption that high prestige prizes confer greater recognition on a prizewinner than low prestige prizes. The results appear unsupportive of the Matthew Effect. On average, for our over 2,000 prizes, we found that a prize's prestige does not predict cumulative advantage. The prizewinners of both high and low prestige prizewinners have similarly sized innovation gaps relative to non-prizewinners when the prize is conferred. Further, we found that the size of the innovation gap does not grow after the prize – the stability of the innovation gap suggests that prizewinning is not associated with the acquisition of undue resources after winning a prize. Finally, inconsistent with Matthew Effects expectations that winning a prize would result in prizewinners acquiring more distinguished coauthors after the prize, we found that prizewinners do not work with coauthors who have greater productivity and impact than the coauthors of non-prizewinners in before or after the prize.

There may be several reasons why our findings do not cohere with some expectations of the Matthew Effect. First, it is conceivable that after being awarded a prize, a scholar may recognize or acquire the skills needed to communicate innovative ideas ([69](#)), making them more likely to publish innovative work in the future without the benefit of an increase in their status. Second, if a prizewinner establishes a new or innovative topic space, the new topic space may give the prizewinner first mover advantages ([70](#)) even if their eminence stays the same. Third, since 1968 when the Matthew effect was first introduced by Robert Merton, it may be that science's self-correcting mechanisms have made prize committees vigilant of avoiding the disadvantages of Matthew effects on scientific reward systems – as some prize committees have endeavored to do ([71](#)).

Nevertheless, these results are positive findings about prizes and reward systems in science. The fact that prizewinners are more innovative than matched non-prizewinners before the prize is in line with the objective that prizes promote innovation. Similarly, the fact that prizewinners after the prize do not seem to receive undue recognition and rewards relative to their actual performance also suggests that prizes on average provide appropriate rewards rather than amplify the acclaim of the already famous.

These results also shed new light on the debate about rates of innovation in science. After a study using the disruption index (33) reported that scientific innovation is declining, policy analysts and scholars have debated the results (72, 73) and called for more theoretical and empirical research on innovation. In contrast to the disruption index, which purportedly measures scientific innovation by the rate at which standalone papers contain revolutionary ideas that topple conventional thought, we used multiple measures that view innovation as a process by which past knowledge is recombined in original or rarely seen before ways. Our findings indicate that different measures of innovation capture different facets and imply different inferences about scientific innovation, suggesting that innovation is a multifaceted construct that requires examination by a range of conceptual and philosophical foundations (74, 75).

Innovation theory should also be advanced. For example, even if the decline in single paper disruptiveness is eventually validated, other possible paths to innovation may exist. For example, incremental but novel papers could result in multiple lines of evidence that together can divide and conquer a larger problem that historically was addressed in a single breakthrough paper. For example, plate tectonics theory synthesized different contributions that tackled a piece of the larger puzzle—continental shift, oceanic structure, magnetic patterns, and earthquake dynamics. Along these lines, if scientists are increasingly equipped to solve a specialized part of a larger puzzle (25), then social and AI technologies that facilitate scientists' abilities to combine ideas and divide and conquer big problems matter to innovation and should be examined. Along these lines, we found that prizewinners have ties that potentially expose them to a broader range of ideas and research collaborations.

Lastly, although this study examines reward systems from the lens of Matthew Effects, achieving equality and eliminating biases in reward systems is an ongoing challenge for future research. Indeed, the tension between the ideal of merit-based scientific recognition and the reality of inequality in prizes goes beyond Matthew Effects and continues to require sustained research. For example, while prizes and innovativeness may be statistically related, other bias may exist in the system in terms of access to education, labs, funding, leadership roles, or even what forms of innovation are chosen to be rewarded. Future studies should continue to improve how scientific contributions are equitably recognized, remove existing bias from scientific reward systems, and endeavor to design systems that fairly reward scientists and promote advancement.

**Acknowledgments:**

The computation in this study was supported by the Center for Computational Science and Engineering of SUSTech, the Northwestern Institute on Complex Systems (NICO), and the Kellogg School of Management.

**Fundings:**
YM is funded by the National Natural Science Foundation of China under Nos. 62006109 and 12031005. CJ is funded by the UKRI Metascience research grants under No. OPP569. BU is funded by the Air Force Office of Scientific Research under Minerva award number FA9550-19-1-0354, the Northwestern University Institute on Complex Systems and Data Analytics (NICO), the Ryan Institute of Complexity, and the Kellogg School of Management at Northwestern University.

**Author contributions:**
Conceptualization: YM, CJ, BU
Methodology: CT, YH, CJ, BU
Investigation: CT
Visualization: CT, YH, BU
Project administration: YM, BU
Supervision: YM, CJ, BU
Writing – original draft, review & editing: CT, YM, CJ, BU

**Competing interests:**
The authors declare that they have no competing interests.

**Data and materials availability:**
The OpenAlex data is publicly available. The data and code for reproducing the main results are shared in this repository: https://github.com/ChaolinTian/innovative-distinctions-of-prizewinners-and-their-networks. For further requests, please email to mayf@sustech.edu.cn.

# Supporting Information for

# "The Innovative Distinctions of Prizewinners and their Networks"

Chaolin Tian[1], Yurui Huang[1], Ching Jin[2], Yifang Ma[1*], Brian Uzzi[3*]

**Affiliations:**

[1]Department of Statistics and Data Science, Southern University of Science and Technology, Shenzhen, Guangdong, China

[2]Centre for Interdisciplinary Methodologies, University of Warwick, Coventry CV4 7AL, United Kingdom

[3]Northwestern Institute on Complex Systems (NICO) and Kellogg School of Management, Northwestern University, Evanston, Illinois, USA

**Emails: correspondence:** mayf@sustech.edu.cn; uzzi@kellogg.northwestern.edu

This PDF file includes:

Supporting text

Figures S1 to S17

Tables S1 to S12

SI References

# CONTENTS

# 1. Data

Our main data sources consist of two main parts: the scientific corpus database which provides the publications, citations, collaborations, and topics for each scientist, and the scientific prizes data. The details are shown below.

## 1.1 Scientific Prize Data

We collected the scientific prize and the list of prizewinners data originally from Wikipedia, Wiki data, and partially the official websites related to prizes in the scientific communities, we manually filled in the missing prizewinning years for the prestigious award, the prestige of an award is based on the wiki page view, which is the average monthly views of each prize's Wikipedia homepage [1]. Since scientists may win multiple prizes in their careers, to eliminate the multiple treatment effects, here we consider the 2,460 prestigious prizes and regard the years of their first prize as the treatment year in our analysis. This results in 7,353 prizewinners whose prizewinning year spans from 1900 to 2018.

Table S1. Scientific Prizes Data Descriptive Table.

|  | Count | Average | SE |
|---|---|---|---|
| **#Prizes** | 2,460 |  |  |
| **#Prizewinner** | 7,353 |  |  |
| **#Recipients per prize** |  | 7.406 | 0.334 |
| **#Prizes per winner** |  | 2.478 | 0.028 |

Table S2. The Number of Scientific Prizewinners, Average Team Size and Innovativeness Measures in Each Discipline.

| Discipline | #Prizewinner | Average Team size | Average Novelty | Average Convergence | Average Interdisciplinarity |
|---|---|---|---|---|---|
| **Art** | 355 | 2.098 (0.009) | 0.025 (0.000) | 0.146 (0.001) | 0.199 (0.000) |
| **Biology** | 764 | 6.333 (0.01) | 0.09 (0.000) | 0.521 (0.001) | 0.188 (0.000) |
| **Business** | 41 | 2.604 (0.023) | 0.055 (0.002) | 0.304 (0.004) | 0.223 (0.001) |
| **Chemistry** | 1,028 | 4.787 (0.005) | 0.132 (0.000) | 0.444 (0.000) | 0.194 (0.000) |
| **Computer science** | 877 | 3.993 (0.01) | 0.062 (0.000) | 0.406 (0.001) | 0.209 (0.000) |
| **Economics** | 232 | 2.449 (0.008) | 0.064 (0.001) | 0.339 (0.001) | 0.229 (0.000) |
| **Engineering** | 33 | 2.693 (0.084) | <0.001 (0.000) | 0.18 (0.014) | 0.212 (0.004) |
| **Environmental science** | 87 | 6.017 (0.034) | 0.21 (0.002) | 0.425 (0.002) | 0.236 (0.000) |
| **Geography** | 131 | 4.808 (0.032) | 0.062 (0.001) | 0.343 (0.002) | 0.199 (0.000) |
| **Geology** | 202 | 4.644 (0.017) | 0.114 (0.001) | 0.349 (0.001) | 0.235 (0.000) |
| **History** | 155 | 1.919 (0.018) | 0.046 (0.001) | 0.185 (0.003) | 0.2 (0.001) |
| **Materials science** | 452 | 5.313 (0.007) | 0.158 (0.001) | 0.421 (0.001) | 0.192 (0.000) |
| **Mathematics** | 643 | 2.342 (0.004) | 0.034 (0.000) | 0.201 (0.001) | 0.149 (0.000) |
| **Medicine** | 420 | 8.065 (0.017) | 0.107 (0.001) | 0.445 (0.001) | 0.185 (0.000) |
| **Philosophy** | 175 | 1.818 (0.01) | 0.031 (0.001) | 0.269 (0.002) | 0.195 (0.000) |
| **Physics** | 1,049 | 6.808 (0.017) | 0.102 (0.000) | 0.418 (0.001) | 0.175 (0.000) |
| **Political science** | 276 | 1.995 (0.009) | 0.022 (0.000) | 0.285 (0.002) | 0.22 (0.000) |
| **Psychology** | 302 | 3.441 (0.011) | 0.085 (0.001) | 0.307 (0.001) | 0.211 (0.000) |
| **Sociology** | 131 | 1.899 (0.018) | 0.008 (0.001) | 0.208 (0.003) | 0.207 (0.001) |

Robust standard errors in parentheses.

## 1.2 Scientific Corpus Database

We used the publicly available database, OpenAlex [2], in our study. OpenAlex integrated multidimensional scientific sources including the Microsoft Academic Graph, the Crossref, PubMed, ORCID, etc. OpenAlex used a sophisticated algorithm to disambiguate authors with the same names using their names, publication records, affiliations, citation patterns, and external identifiers like ORCID. As of July 2022, the OpenAlex snapshot (we used in our study) has included about 213M authors, over 240M works, and about 109,000 institutions. Papers are tagged with multilevel fields of study classifications using NLP and verified the field concept terms with Wikipedia, the highest level (level-0) of classification contains 19 disciplines, i.e., Political science, Philosophy, Economics, Business, Psychology, Mathematics, Medicine, Biology, Computer science, Geology, Chemistry, Art, Sociology, Engineering, Geography, History, Materials science, Physics, Environmental science.

We used the names, affiliations, and research fields to match each prizewinner to the OpenAlex. We identified each prizewinner's research discipline as the most frequent level-0 concept according to all her/his publications. In our study, we further extract the yearly number of publications, the yearly number of citations, the first publication year, the total number of publications, and the total number of citations for all scholars with at least 10 publication records, which contain over 6M authors.

## 2. Innovativeness Measures and Validation

## 2.1 Novelty

Novelty is used to quantify how the work leverages past knowledge to produce new knowledge by looking at the journal combinations of the referenced papers [3], which captures the rarity of knowledge combinations. Following prior research, we measure novelty for each paper by examining the occurrence of journal combination pairs of prior work referenced in the paper, compared to the expected occurrence of journal pairs when the reference list is shuffled randomly while keeping the publication years and the number of citations of each reference. We generated a z-score for each journal pair, then each paper has a z-score distribution from all combinations of its references list. The z-score is calculated by comparing the observed frequency of journal pairs that appear within a paper's reference list and the expected distribution of journal pairs created by randomized citation networks. Details of the calculation process are shown in Figure S1(A). We calculated the novelty measure for all journal and conference publications from the OpenAlex and validated the main result related to the novelty measure from the work of Uzzi *et al.* [3], that is, as shown in Figure S1(C), the papers with both novel and conventional properties are more likely to be hit paper (top 5% citation).

In our study, we defined a paper as a novel paper if the paper combined high median conventionality and high tail novel combination, where high median conventionality indicates the paper's median z-score is in the upper half of all median z-scores in that year and high tail novel combination indicates the paper's $10^{th}$ percentile z-score is below zero. The definition is as below:

$$\text{Novelty} = \begin{cases} 1, & \text{median z-score>all median z-scores \& } 10^{th} \text{ z-score} \leq 0; \\ 0, & \text{otherwise.} \end{cases} \quad (1)$$

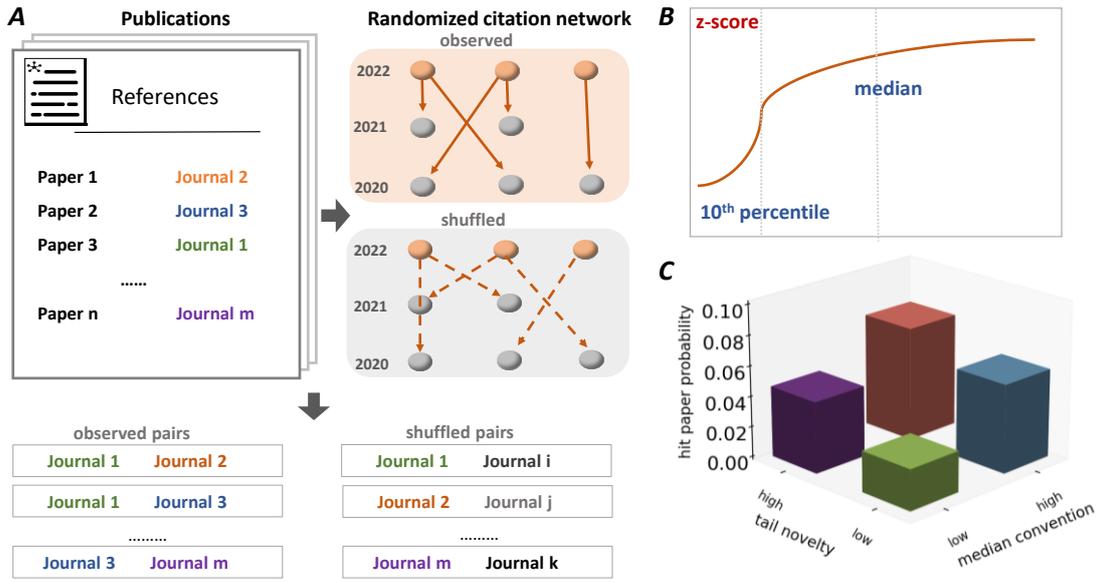

**Figure S1. Novelty Measure Based on Referenced Prior Knowledge Combinations.** (A) shows the detailed calculation process. We first counted the population-wide observed frequency of each journal pair for all referenced pairs from a given publication year. Second, we randomly shuffled the citation links between papers while keeping the publication years and the number of citations of each reference, after that, we counted the shuffled frequency of each journal pair for all referenced pairs. For each paper, we obtained its referenced journal pairs' z-score distribution by calculating each journal pair's z-score: $z = (x - \bar{x})/s$, where $x$ is the observed journal pairs' frequency, $\bar{x}$ is the mean and $s$ is the standard deviation of the number of journal pairs obtained from the randomization. (B) shows the z-score distribution of focal paper. A paper is novel if the paper combines a high median conventionality (median z-score greater than all papers' median z-scores in the same publication year) and a high tail novel (10th z-score below zero) combination. (C) shows our reproduced result, which is consistent with the work of Uzzi *et al.* [3], that is, the papers with both novel and conventional properties are more likely to be a hit paper (top 5% citation).

## 2.2 Convergence

Convergence is used to reveal how a work is following the frontiers of science by looking at the age distributions of its referenced works [4]. We measure the convergence of a paper by examining the age distribution of its references, denoted as $D$, which includes the age differences between a paper's publication year and the publication years of its references. A paper is convergence if its age difference distribution exhibits a low mean and large coefficient of variation (CV) compared to the global yearly mean and CV. Details of the method are shown in Figure S2(A).

According to the definition, a paper is a convergent paper if it satisfies:

$$\text{Convergence} = \begin{cases} 1, D_\mu < D_{\mu^*}, D_\theta > D_{\theta^*}; \\ 0, \text{ else.} \end{cases} \qquad (2)$$

where $D_\mu$ and $D_\theta$ are the mean and CV of a focal paper's age distribution. $D_{\mu^*}$ and $D_{\theta^*}$ are the corresponding yearly global mean and CV, computed based on reference age distributions for all papers published in the same year as the focal paper. We validated that the paper of convergence has a higher probability of being hit paper, which is consistent with Mukherjee et al.'s results [4] (see Figure S2 (B)).

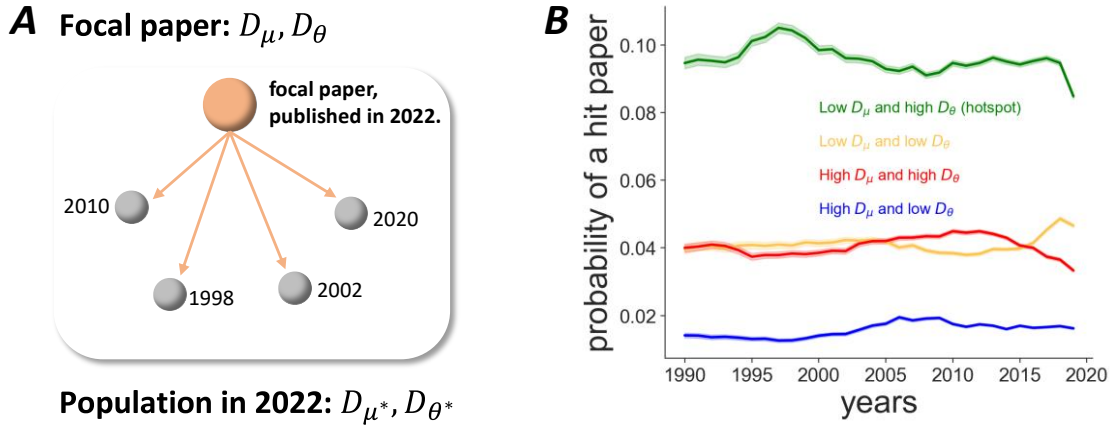

**Figure S2. Convergence Measure Based on the Reference Age Distribution.** (A) shows the calculation process. For example, the focal paper's references' age differences distribution is 12, 24, 20, and 2. So $D_\mu = 14.5$, $D_\theta = \sigma/D_\mu = 0.58$, where $\sigma$ is the standard deviation of the age distribution. Then we compared $D_\mu$, $D_\theta$ with the global yearly age distribution mean ($D_{\mu^*}$) and CV ($D_{\theta^*}$). A paper is convergent if its age difference distribution exhibits a lower mean and large CV. (B) shows that papers with low mean and large CV have higher probabilities of being hit papers, which is consistent with Mukherjee, et al.'s results [4].

## 2.3 Interdisciplinarity

Interdisciplinarity reflects the extent a paper's research content fills in a multi-disciplinary domain [5, 6], by checking the broadness of subject categories associated with a paper's reference or citations. Here we quantify the interdisciplinarity of a focal paper based on the formula: $\Delta = \sum_{ij(i\neq j)} d_{ij} p_i p_j$, where $p_i$ and $p_j$ are the proportion of the focal paper's citing papers in subject category $i$ and $j$. $d_{ij}$ is the cosine dissimilarity score between $i$ and $j$ which is calculated through the co-citation matrix for all papers [6]. The value $\Delta$ ranges from 0 to 1, and higher values indicate that the paper is more interdisciplinary.

We used level-1 concepts in OpenAlex as subject categories in our analysis. In the OpenAlex taxonomy, each paper is tagged with one or more concepts organized in a hierarchical structure. The level-1 concepts represent the broadest fields of study and include a total of 284 categories. Details of how these categories were used in the analysis are provided in Figure S3.

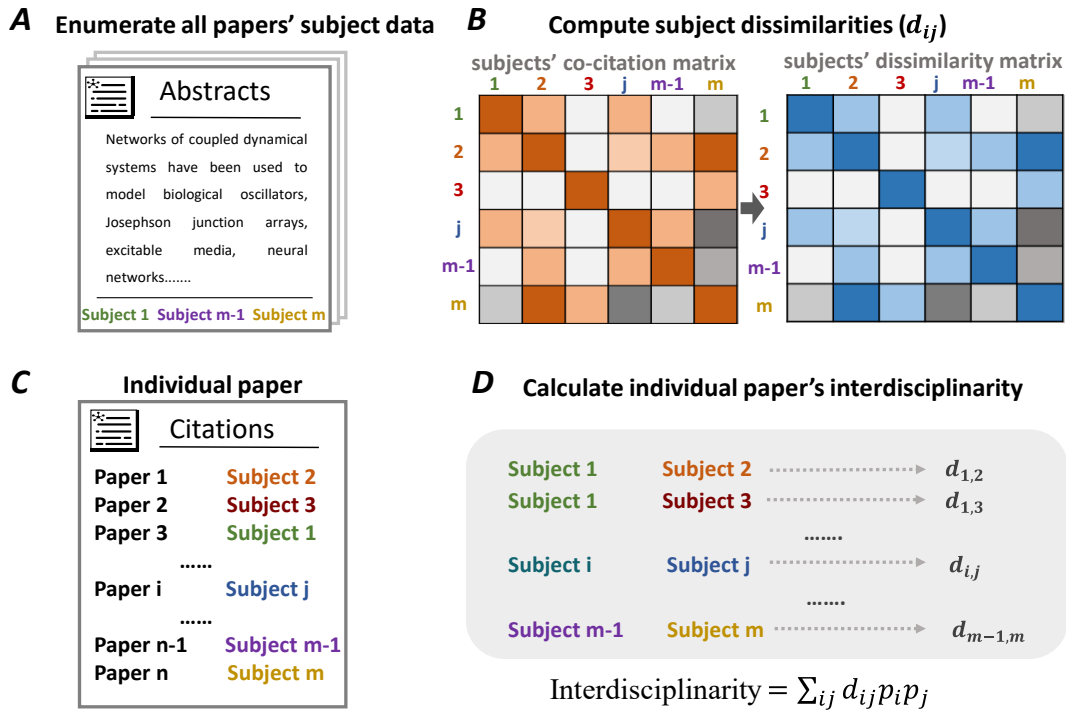

Interdisciplinarity $= \sum_{ij} d_{ij} p_i p_j$

**Figure S3. Interdisciplinarity Measure Based on Subject Pairings.** (A) We used level-1 concepts from OpenAlex as subject categories and computed the co-citation frequency for each subject pair based on all citing papers in the OpenAlex database. This yielded a subject co-citation matrix, where each cell represents the number of times a subject pair was co-cited. (B) We converted the co-citation matrix into a cosine dissimilarity matrix, defined as: $d_{ij} = 1 - \left( \sum_k n_{ik} n_{jk} \big/ \sqrt{\sum_k n_{ik}^2 \sum_k n_{jk}^2} \right), \ (k = 1, \dots, m)$, where $n_{ik}$ is the number of times subjects $i$ and $k$ were co-cited across all papers. (C) For each focal paper, we identified all subject category pairs among its citing papers. (D) We then computed the Stirling index [6]: $\Delta = \sum_{ij(i \neq j)} d_{ij} p_i p_j$, where $p_i$ and $p_j$ are the proportion of the focal paper's citing papers in subject category i and j, and $d_{ij}$ is the cosine dissimilarity between categories $i$ and $j$, as defined in (B).

## 2.4 Correlation and Evolution Trend of Innovativeness Measures

To examine whether the three innovativeness measures used in our study—novelty, convergence, and interdisciplinarity—capture distinct dimensions of scientific innovativeness, we conducted two complementary analyses.

First, we computed pairwise Pearson correlation coefficients among the three indicators (see Table S3). The results show that the correlations are consistently low, with coefficients ranging from 0.018 to 0.037, although all are statistically significant due to the large sample size. This suggests that the three measures are weak correlated and likely reflect different aspects of innovativeness.

Second, we conducted a principal component analysis (PCA) to further explore the relationships among the three measures (See Figure S4)[7]. The PCA loading plot visualizes how each indicator contributes to the first two principal components. Novelty and convergence load strongly but in opposite directions along the first principal component, whereas interdisciplinarity aligns along a distinct axis closer to the second component. The directions and magnitudes of the loadings indicate that the three indicators may represent diverse dimensions of innovativeness.

These findings support the validity of using all three measures in our analysis to capture the multifaceted nature of scientific innovation.

**Table S3 The Pearson Correlation Coefficients for the Three Innovativeness Measures.**

|  | **Novelty** | **Convergence** | **Interdisciplinarity** |
|---|---|---|---|
| **Novelty** | 1 |  |  |
| **Convergence** | 0.018*** | 1 |  |
| **Interdisciplinarity** | 0.035*** | 0.037*** | 1 |

***: p<0.001 (Pearson correlation coefficient test).

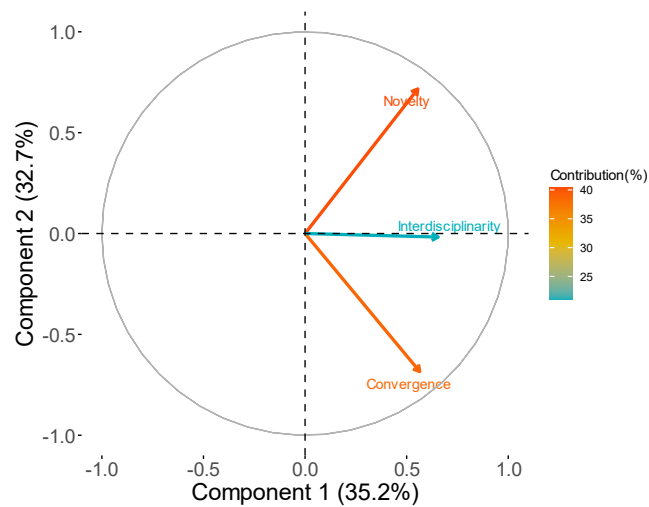

**Figure S4. Principal Component Analysis of Three Innovativeness Measures Based on All Papers in OpenAlex.** This PCA loading plot illustrates the relationships among three innovativeness measures—novelty, convergence, and interdisciplinarity—calculated from the full set of scientific papers in the OpenAlex database. The plot shows the loadings of these indicators on the first two principal components. Arrows represent the direction and magnitude of each indicator's contribution to the component space, with colors indicating the relative strength of loading (warmer

tones denote stronger associations).

To determine whether the innovation trajectories of prizewinners represent idiosyncratic patterns, we compared the long-term trends in novelty, convergence, and interdisciplinarity between prizewinners and the overall scientific population (based all data in OpenAlex).

As shown in Figure S5, prizewinners (orange lines) consistently exhibit higher average levels across all three innovativeness measures compared to the global scientific baseline (blue lines). However, despite these elevated levels, their temporal dynamics of change closely mirror global trends across disciplines and measures. Specifically, both groups exhibit rising interdisciplinarity over time, parallel shifts in convergence, and similar non-linear patterns in novelty. This alignment in both direction and shape of the trends suggests that while prizewinners lead in innovation, they remain embedded within the broader systemic trajectory of scientific development.

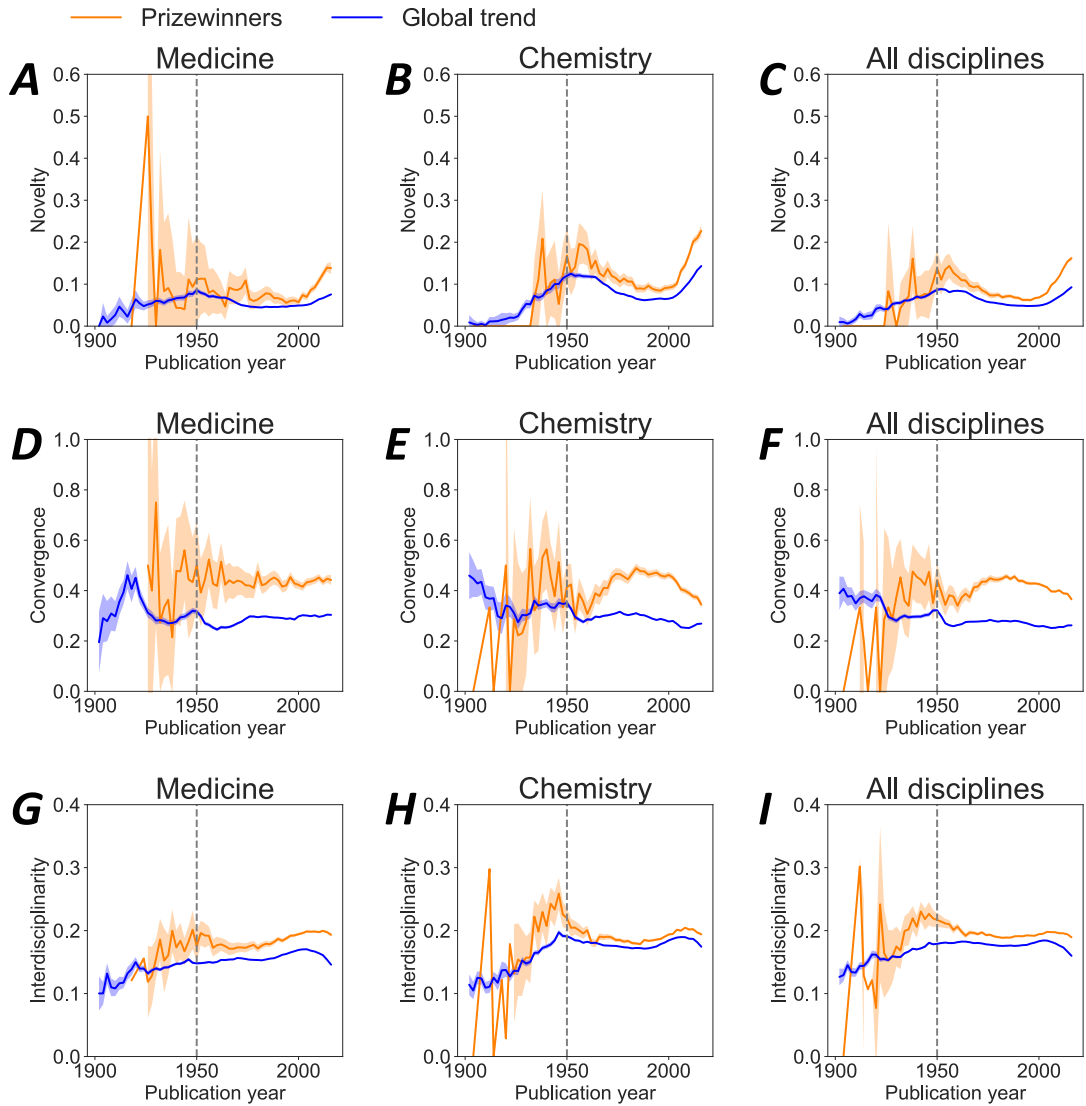

**Figure S5. Long-Term Trends in Novelty, Convergence, and Interdisciplinarity Among Prizewinners Compared to Global Trends.** (A–C) show average novelty over publication years for prizewinners (orange) and the global scientific baseline (blue), respectively for Medicine, Chemistry, and all disciplines. (D–F) show convergence, and (G–I) show interdisciplinarity, following the same layout. The vertical dashed line denotes the average award year for the prizewinner sample. Shaded areas represent 95% confidence intervals.

## 3. Matching Designs

We used matching methods to reduce the effects of confounders when estimating the treatment effects and when the natural experiments are impossible, the method will match the indistinguishable pair of observations that have close characteristics before the treatment. In this work, we used both coarsened exact matching (CEM) and dynamic optimal matching (DOM) [8, 9] to create equivalent prizewinners and control groups. The CEM conducts the exact matching based on variables that are coarsened into groups, which balances the treated and control groups in both categorical and continuous variables. DOM matches individuals on a set of time-varying characteristics to ensure that their fixed attributes and overtime behavior changes are aligned with an array of stringent statistical tests to measure their equivalence.

We combined the benefits of CEM and DOM in our matching designs. The CEM was used to match the non-prizewinners from millions of scientists (each has clear publication, year, and discipline records) in our candidate pool with coarsened categorical characteristics, we matched 200 non-prizewinners for each prizewinner. Then, within each 1:200 group, we use the DOM to further select the non-prizewinners to make sure that their time-varying characteristics are close enough. Figure S6 illustrates the matching procedures.

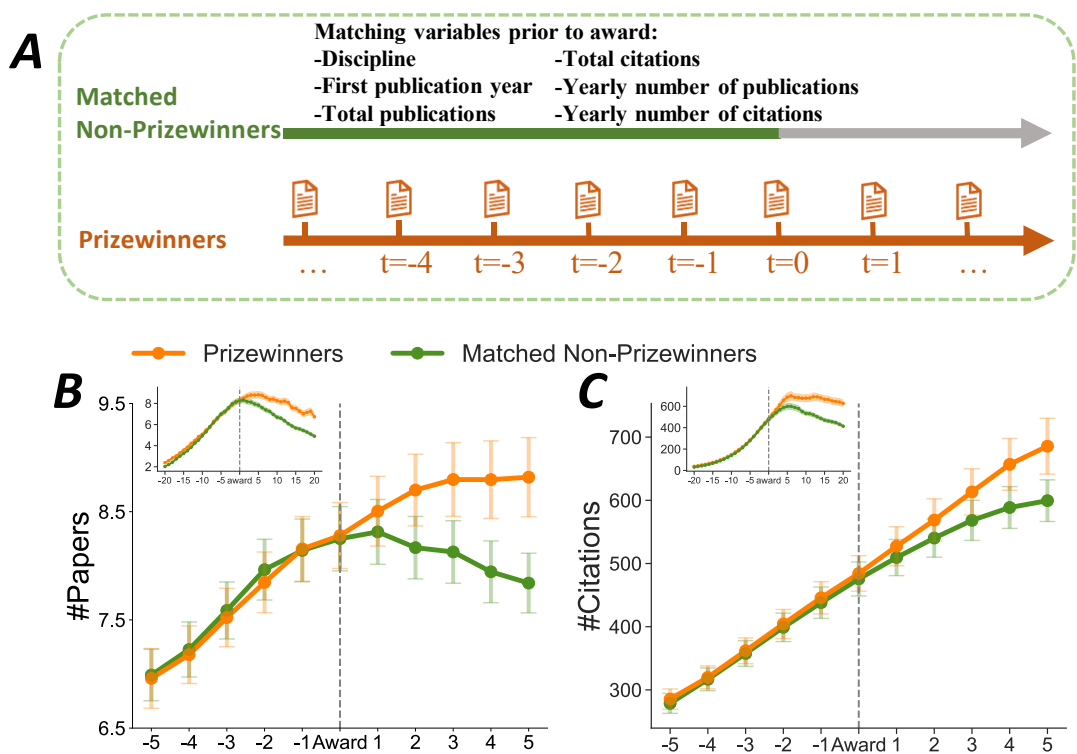

**Figure S6. Matching Design and Validation.** (A) Procedures for matching the non-prizewinner group, each prizewinner is to a non-prizewinner who is indistinguishable in the matching variables before the prizewinner's prizewinning year by combining the CEM and DOM techniques. (B) and (C) show the number of papers and the number of citations for the prizewinners (PWs) and matched non-prizewinners (NPWs). Error bars show the 95% confidence intervals. Insets show the long-term trends within 20 years before and after the prizewinning.

## 3.1 Two Step Matching

### 3.1.1 Matching Procedures

### Step 1: Coarsened Exact Matching

For each prizewinner, we identify his/her first publication year, first prizewinning year, research discipline, total number of publications, and the total number of citations, for each prizewinner, we matched a set of non-prizewinners (up to 200) who have:

- the same research discipline as the prizewinner, here we used the level-0 concept in OpenAlex to tag each scientist's discipline according to their publications.

- the similar research age, i.e., the career starting year (year of the first publication) is within 2 years before or after the prizewinner's career starting year.

- the total number of publications, which has a maximum 30% fluctuation above or below the prizewinners before the prizewinning year.

- the total number of citations, which has a maximum 30% fluctuation above or below the prizewinners before the prizewinning year.

### Step 2: Dynamic Optimal Matching

Based on Step 1, we further match the yearly number of publications and yearly number of citations dynamically. We used two different distance measures to conduct the matching.

At first, we define the distance measure ($d_{i,j}^p$) for prizewinner $i$ and the non-prizewinner $j$ based on the yearly number of publications (the same procedure for the yearly number of citations, $d_{i,j}^c$) 5 years before the prizewinning year, as follows,

$$d_{i,j}^p = \frac{1}{t_0 + 1} \sum_{t=t^*-t_0}^{t^*} \frac{|Y_i(t) - Y_j(t)|}{Y_i(t)}, \qquad (3)$$

where $Y_i(t)$ denote the publication count of prizewinner $i$ in year $t$, $t^*$ indicates the prizewinning year of prizewinner $i$, and we traced back the matching for $t_0 = 5$ years before the prizewinning year. The same approach applies to calculating the citation distance $d_{i,j}^c$, using citation counts instead of publications.

Due to computational cost and the fact that publication and citation increase with time,

the recent years close to the prizewinning year are more necessary to control, here we define $t_0 = 5$ to capture a 6-year window leading up to the prize year, including the prizewinning year itself. For the DOM we only matched 6 years. But after we matched the 6 years, we found the citations and pubs before 6 years are automatically closed enough.

To ensure comparability and balance between the prizewinners and non-prizewinners across the entire system, we matched each prizewinner with up to 5 scientists from candidate pool. The matching was based on minimizing the distance between prizewinners and candidates, using thresholds of $|d_{i,j}^p| \leq 0.6$ for publication distance and $|d_{i,j}^c| \leq 0.6$ for citation distance. Only non-prizewinners meeting both criteria were retained; otherwise, the group was discarded. This threshold ensured sufficient pre-award similarity while maximizing the number of PWs that could be successfully matched, resulting in 4,470 matched groups of prizewinners (PWs) and non-prizewinners (NPWs). Robustness checks using alternative thresholds are presented in Section 3.1.2 and Figures S7–S8.

### 3.1.2 Robustness Check under Different Thresholds

In the matching procedure, our default threshold was set to 0.6 (i.e., $|d_{i,j}^p| \leq 0.6$ for publication and $|d_{i,j}^c| \leq 0.6$ for citation), which preserves a balance between ensuring close pre-treatment comparability and maximizing the coverage of prizewinners who could be matched (at least 80% of prizewinners are matched). At this threshold level, we were able to retain the largest number of prizewinners while still ensuring that the matched NPWs were sufficiently similar in terms of their scientific trajectories. Stricter thresholds (e.g., 0.3 or 0.4) eliminate a larger portion of prizewinners due to unmatched candidates, whereas looser thresholds would risk introducing mismatches.

To assess the robustness of our results to different threshold choices, we conducted the same matching procedure under three alternative threshold levels: 0.3, 0.4, and 0.5. Figures S7 and S8 display the comparison of matched PWs and NPWs in terms of their publication and citation trends (Figure S7) and their innovation trajectories across all three measures (Figure S8), using these different thresholds.

The results consistently show that PWs exhibit significantly higher post-award productivity and innovativeness than their matched NPWs, and that the overall trends and treatment effects are stable across different matching thresholds. This suggests that our findings are not driven by a particular distance cutoff and remain robust across reasonable parameter choices.

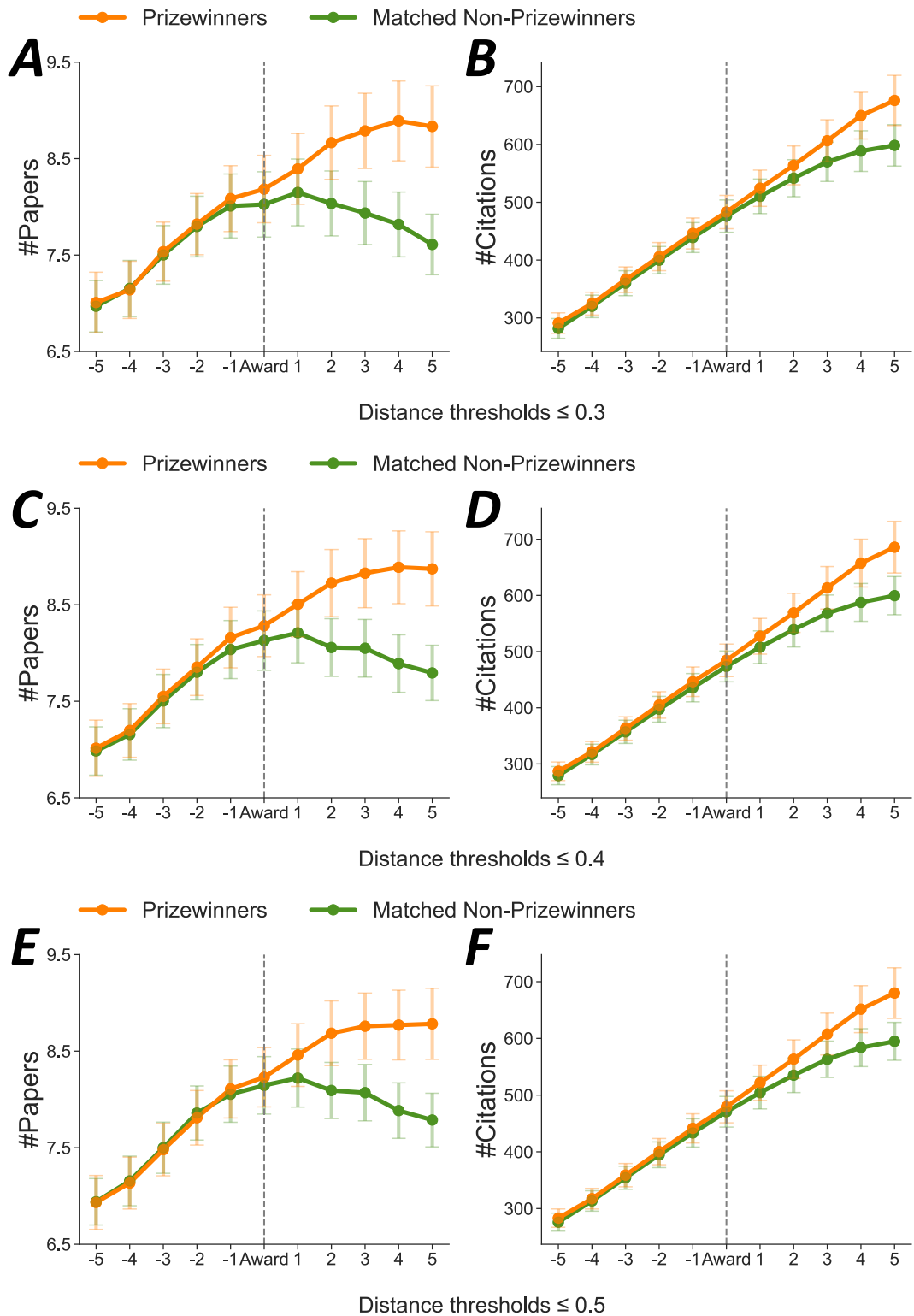

**Figure S7. Scientific Output and Citation Trends Before and After Prizewinning for PWs and NPWs.** (A), (C), and (E) show the average number of papers (#Papers) per year; (B), (D), and (F) show the average number of citations (#Citations) per year. Orange lines represent prizewinners, and green lines represent their matched non- prizewinners. The vertical dashed line at year 0 denotes the prizewinning year. Each panel corresponds to a different matching threshold based on dynamic distance: (A–B) use a threshold of 0.3, (C–D) use 0.4, and (E–F) use 0.5. Error bars represent 95% confidence intervals.

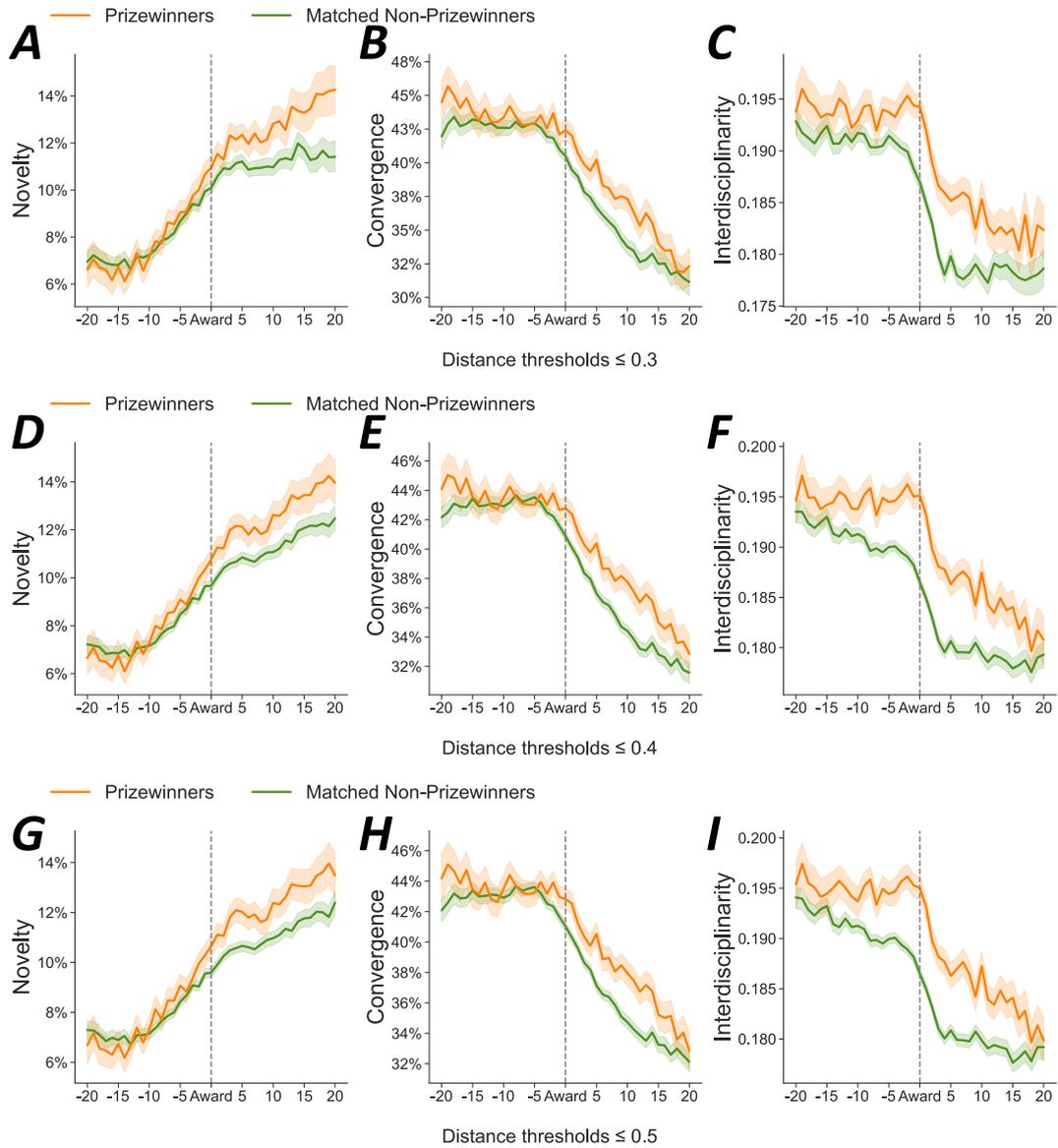

**Figure S8. Innovation Dynamics of PWs and NPWs by Different Thresholds.** (A), (D), and (G) show the average novelty of paper; (B), (E), and (H) show convergence; (C), (F), and (I) show interdisciplinarity. Orange lines represent prizewinners, and green lines represent matched non-prizewinners. The vertical dashed line at year 0 indicates the prizewinning year. Each row corresponds to a different dynamic matching threshold: (A–C) use a threshold of 0.3, (D–F) use 0.4, and (G–I) use 0.5. Shaded areas represent 95% confidence intervals.

## 3.2 Robustness checks

### 3.2.1 Robustness Check by Matching with Team Size as a Covariate

To examine the robustness of our results to differences in team size, we extended the baseline matching procedure by incorporating average team size—measured as the mean number of coauthors per paper prior to the prizewinning year—as an additional matching variable. In this specification, we added Step 1 (CEM) of the two-step procedure (see Section 3.1.1) to include only candidates whose average team size was within ±30% of the corresponding prizewinner's pre-award average. All other matching criteria in Step 1 (discipline, career age, total publications, and total citations) were retained.

The subsequent Step 2 (DOM) based on yearly publications and citations—and the distance threshold ($\leqslant 0.6$) remained unchanged. This ensures that matched controls are not only similar in career trajectory and productivity but also comparable in collaboration scale before the treatment.

As shown in Figure S9, the observed divergence in productivity and innovativeness measures between prizewinners and matched non-prizewinners persists even after adjusting for team size, reaffirming the robustness of our core findings.

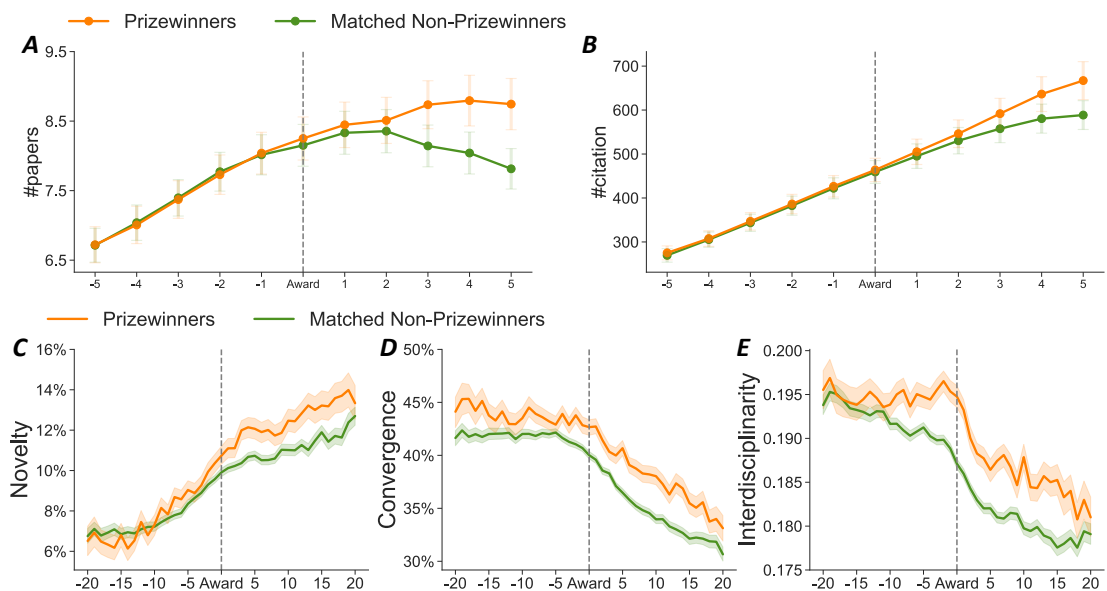

**Figure S9. Innovation Dynamics of PWs and NPWs After Additionally Matching on Team Size.** (A-B) show the average number of papers and citations per year for prizewinners (orange) and matched non-prizewinners (green), respectively. (C–E) show the corresponding trends in novelty, convergence, and interdisciplinarity. Matching procedures include covariates for research discipline, career age, productivity, citation impact, and now also team size. The vertical dashed line at year 0 denotes the prizewinning year. Shaded areas represent 95% confidence intervals.

### 3.2.2 Robustness Check Using an Alternative Distance Metric

Furthermore, we deployed an optional matching by using a new distance measure and applied the DOM within each group to ensure the balance within each group. We define the distance measure ($\theta_{i,j}$) for PW $i$ and the candidate $j$ based on the yearly number of publications and yearly number of citations:

$$\theta_{i,j} = \frac{\sum_{n=1}^{N} \sum_{t=t^*-t_0}^{t^*} \left( \log Y_{i,n}(t) - \log Y_{j,n}(t) \right)^2}{N * (t_0 + 1)}, \tag{4}$$

where $Y_{i,n}$ represents the quantity for PW $i$. $N = 2$ is the matched categories (i.e., $n = 1$ for *publications*, 2 for *citations*). $t$ measures the number of years before the prizewinning year. $t^*$ represents the prizewinning year for PW $i$, and $t_0 = 5$ capturing a 6-year span, including the prizewinning year. Using this distance $\theta_{i,j}$, we identified up to 40 close-distance scientists for each PW.

Then we matched each prizewinner with corresponding candidates within each group by minimizing their covariate distances using optimal matching based on the Mahalanobis distance. This optimization was implemented via the R package "MatchIt" [10], which applies a network flow-based algorithm to identify globally optimal pairings [8, 11]. The Mahalanobis distance between PW $i$ and candidate $j$ is calculated as follows,

$$md_{ij} = \sqrt{(\vec{x}_i - \vec{x}_j) S^{-1} (\vec{x}_i - \vec{x}_j)'}, \tag{5}$$

where $\vec{x}$ is the covariates (yearly #*publications* and #*citations*) vector with the dimension of 12, and $S$ is a scaling matrix, typically the covariate matrix of the covariates. It is not sensitive to the scale of the covariates and accounts for redundancy between them [12]. Furthermore, we replicated our main analysis, which yielded consistent results, reinforcing the robustness of our findings (see Figure S10).

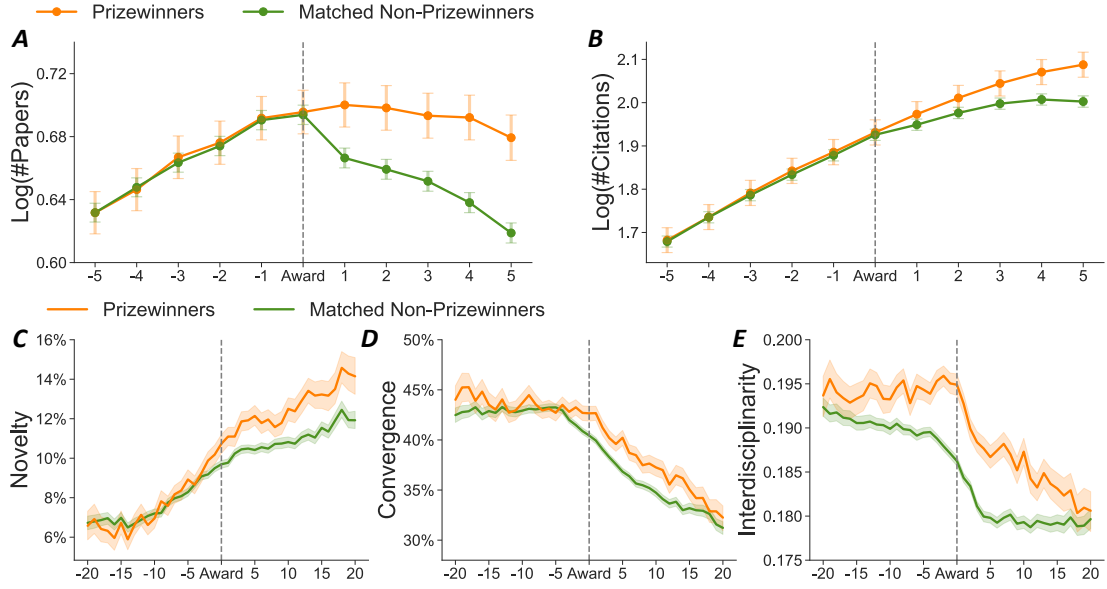

**Figure S10. Innovation Dynamics of PWs and NPWs for the Matching Method with Distance $md_{ij}$.** (A) shows the number of papers and the number of citations (B) by year for the PWs and NPWs from the person match group. For the three innovativeness measures and the three groups of scientists. The average of novelty (C), convergence (D), and interdisciplinarity (E) are shown. Colors represent different groups: the PWs group (orange lines), the NPWs group (green lines). Areas under curves are the 95% CIs.

### 3.2.3 Assessing Matching Quality under Different Distance Metrics

We employed *t*-tests to verify that there was no statistically significant difference between the prizewinner group and matched non-prizewinner group in terms of the yearly number of publications and citations within the 5-year match period, Table S4 shows that all variables have *p*-values greater than 0.5. Further, we used Standardized Mean Differences (SMD) to test the balance of the covariates after matching. An SMD value less than 0.1 indicates a balanced match between the prizewinner and matched non-prizewinner groups across matching variables. The SMD is defined as follows:

$$SMD = \frac{\bar{x}_i - \bar{x}_j}{\sqrt{\left(S_i^2 + S_j^2\right)/2}},$$ (6)

where $\bar{x}$ represents the covariates mean of the prizewinners' group and matched non-prizewinners' group. $S^2$ represents the variance for each group. Table S4 demonstrates that all SMD values are less than 0.1.

**Table S4. Balance and Closeness Test for Matchings with Different Distance Measures.** SMD<0.1 means the covariates are balanced. P-value>0.5 indicates that there is no statistically significant difference between the

| t | Publication ($d_{i,j}$) | | Publication ($md_{i,j}$) | | Citation ($d_{i,j}$) | | Citation ($md_{i,j}$) | |
|---|---|---|---|---|---|---|---|---|
| | SMD | p-value | SMD | p-value | SMD | p-value | SMD | p-value |
| -5 | -0.0036 | 0.8664 (-0.1682) | -0.0001 | 0.9950 (-0.0062) | 0.0123 | 0.5599 (0.5829) | 0.0034 | 0.8380 (0.2045) |
| -4 | -0.0055 | 0.7963 (-0.2581) | -0.0032 | 0.8442 (-0.1965) | 0.0051 | 0.8102 (0.2402) | 0.0005 | 0.9764 (0.0296) |
| -3 | -0.0072 | 0.7332 (-0.3409) | 0.0074 | 0.6504 (0.4532) | 0.0063 | 0.7663 (0.2972) | 0.0052 | 0.7521 (0.3159) |
| -2 | -0.0123 | 0.5601 (-0.5827) | 0.0044 | 0.7861 (0.2714) | 0.0068 | 0.7483 (0.3209) | 0.0087 | 0.5937 (0.5334) |
| -1 | 0.0014 | 0.9477 (0.0656) | 0.0026 | 0.8739 (0.1587) | 0.0090 | 0.6714 (0.4243) | 0.0074 | 0.6525 (0.4503) |
| 0 | 0.0029 | 0.8927 (0.1349) | 0.0037 | 0.8204 (0.2271) | 0.0093 | 0.6614 (0.4381) | 0.0056 | 0.7331 (0.341) |

SMD: standardized mean differences. The t-values are in parentheses.

### 3.2.4 Robustness Check with Dynamic Matching on Innovativeness

To further validate the robustness of our findings, we conducted an alternative matching procedure in which non-prizewinners (NPWs) were further filtered based on their similarity to prizewinners (PWs) in innovation dynamics prior to the prize year.

In Step 1, we retained the same baseline matching criteria as described in Section 3.1.1—matching on discipline, career age, total publications, and total citations—while additionally controlling for average team size before the prizewinning year (within a ±30% range). This step ensured comparability in both research productivity and collaboration scale.

In Step 2, we replaced the dynamic matching on yearly publication and citation with a distance-based comparison of innovativeness measures. Due to the discontinuity by years (a scholar may not publish paper in a specific year), for each innovativeness measure—novelty, convergence, and interdisciplinarity—we calculated the average value within four consecutive 5-year windows covering the 20 years preceding the prize year: $[-20, -15), [-15, -10), [-10, -5), [-5, 0)$. This yielded a 4-dimensional innovation vector for each PW and his/her corresponding candidates (derived from step CEM), based on a given innovativeness measure $k$.

We then computed the Euclidean distance between PW $i$ and candidate $j$ as:

$$d_{i,j}^{(k)} = \sqrt{\sum_{\Delta \in \mathcal{D}} (\bar{I}_{i,\Delta}^{(k)} - \bar{I}_{j,\Delta}^{(k)})^2}, \tag{7}$$

where $k \in \{\text{Novelty, Convergence, Interdisciplinarity}\}$ denotes the innovativeness measures; $\mathcal{D} \in \{-20,-15,-10,-5\}$ denotes the set of pre-prizewinning periods; $\bar{I}_{i,\Delta}^{(k)}$ is the 5-year average of indicator $k$ for PW $i$ in period $\Delta$; $\bar{I}_{j,\Delta}^{(k)}$ is the corresponding value for candidate $j$; $d_{i,j}^{(k)}$ is the Euclidean distance based on metric $k$.

For each prizewinner, we selected up to five candidates as the matched non-prizewinners (NPWs) with the smallest distances under the constraint that $d_{i,j}^{(k)} \leq 0.1$. This ensures close similarity in pre-award innovation patterns. These alternative matched samples, which control for pre-award innovation dynamics provide additional support for the robustness of our core findings.

Figure S11 illustrates the results from this matching design: in all three innovativeness measures, PWs show a higher innovation following the award compared to their NPWs, consistent with our main conclusion.

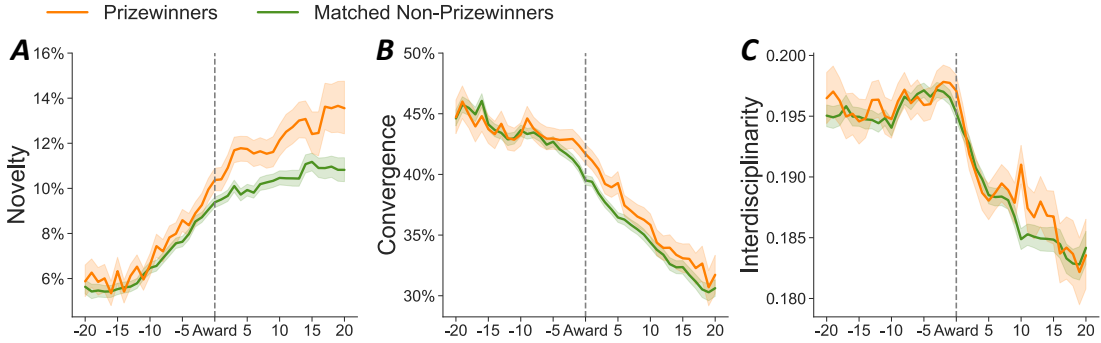

**Figure S11. Innovation Dynamics of PWs and NPWs Based on Innovativeness Matching.** Each panel shows the average values of (A) novelty, (B) convergence, and (C) interdisciplinarity from 20 years before to 20 years after the prizewinning year (denoted by the dashed vertical line). Prizewinners (orange) are matched to non-prizewinners (green) based on pre-award similarity in each respective innovativeness measure using a four-dimensional Euclidean distance over 5-year averages in the [-20, -5] period. Across all three measures, prizewinners exhibit significantly higher post-award innovativeness than their matched counterparts, indicating that the observed innovation advantage is robust even when matching on pre-award innovation dynamics. Error bars represent 95% confidence intervals.

### 3.2.5 Robustness Check by Excluding Review Papers

To ensure that our findings on scientific innovativeness are not driven by review papers—which typically summarize existing knowledge rather than contributing original ideas—we conducted robustness checks by excluding all review papers from the analysis.

Figure S12 shows the trajectories of the three innovativeness measures for PWs and their NPWs after removing review papers. The observed patterns remain consistent with our main findings. These patterns confirm that the innovation advantage of prizewinners is not an artifact of review article production.

Furthermore, we re-estimated the fixed-effects OLS regressions using only original research articles. As shown in Table S10, the positive association between prizewinning and innovativeness remains statistically significant after controlling for team size, matched groups, author position, and publication year. These results reinforce the robustness of our conclusions.

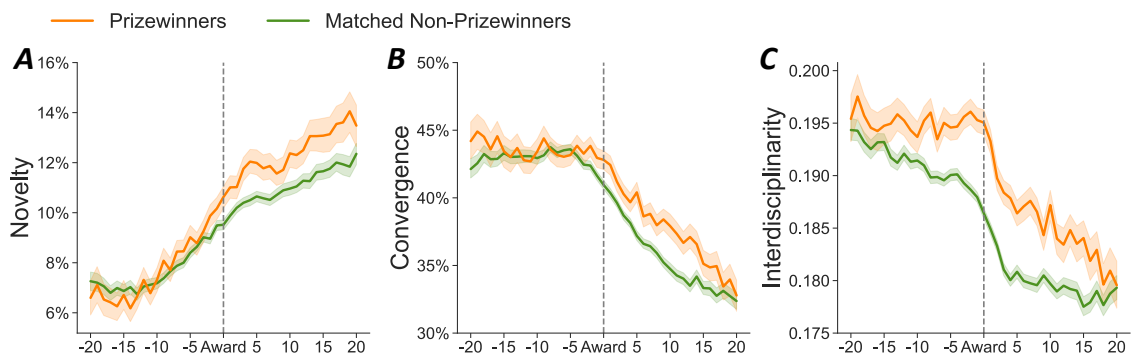

**Figure S12. Innovation Dynamics of PWs and NPWs After Excluding Review Papers.** (A–C) show the trajectories of novelty, convergence, and interdisciplinarity, respectively, for prizewinners (orange) and matched non-prizewinners (green) over time relative to the award year. Shaded areas represent 95% confidence intervals.

### 3.2.6 Robustness Check across Prize Prestige

To examine whether the observed effects vary with the prestige of awards, we stratified the prizewinners in our matched sample (as constructed in Section 3.1.1) into two groups based on the prestige of their first prize. Following our main definition (Section 1.1), we classified a prize as top-tier if its Wikipedia page view count ranked in the top 25% among all prizes, and as non-top otherwise.

We then compared the innovation trajectories of top PWs and non-top PWs against NPWs, using the same matched pairs identified through our two-step procedure. Importantly, this analysis involved no changes to the matching design; only the prizewinners were split by prize prestige to assess heterogeneity in treatment effects.

As shown in Figure S13, both top and non-top PWs exhibit a clear post-award increase in novelty, convergence, and interdisciplinarity, relative to their matched NPWs. The trends are consistent across the two groups, indicating that the observed innovation advantage of prizewinners is not solely driven by the recipients of the most prestigious prizes.

These results demonstrate that our main findings hold across a broad range of award types and are robust to heterogeneity in prize prestige.

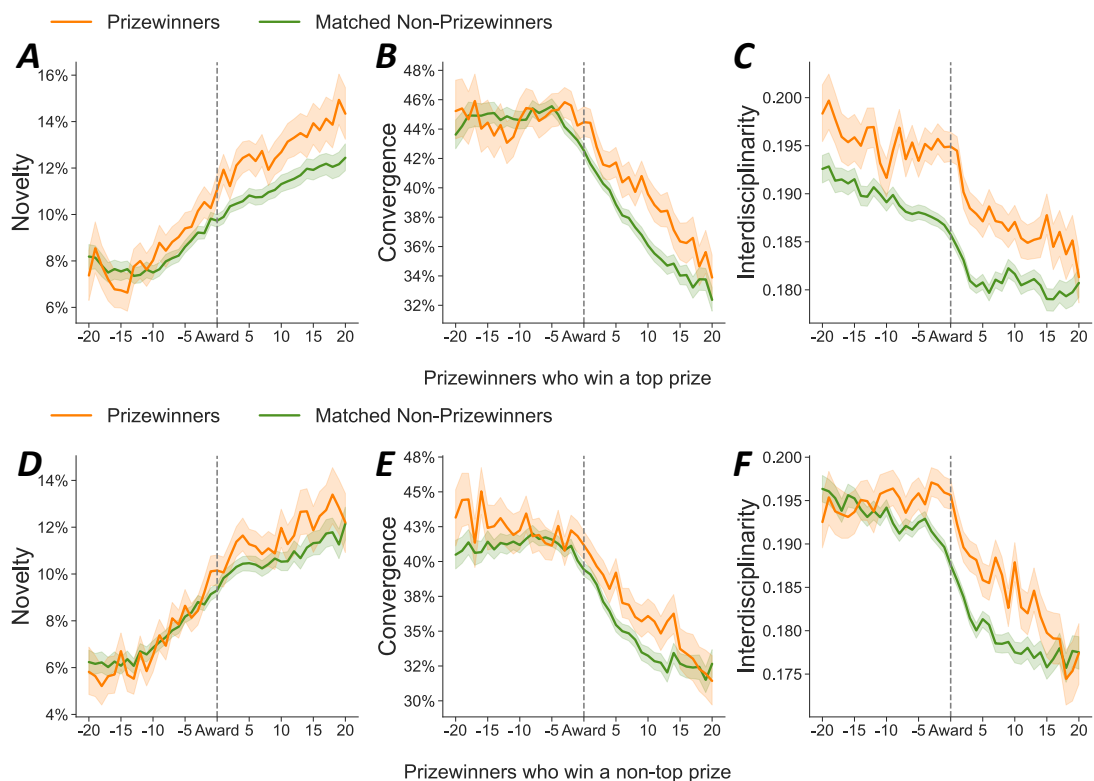

**Figure S13. Innovation Dynamics of PWs and NPWs by Different Prize Prestige.** Changes in novelty, convergence, and interdisciplinarity before and after award receipt, by prize prestige. (A–C) present trends for prizewinners who receive a top prize, and (D–F) show results for those who receive a non-top prize. In each row, panels show the average values of (A/D) novelty, (B/E) convergence, and (C/F) interdisciplinarity. Orange lines represent prizewinners, and green lines represent their matched non- prizewinners. The vertical dashed line at year 0 indicates the prizewinning year. Shaded areas represent 95% confidence intervals.

### 3.2.7 Robustness Check across Disciplines and Time Periods

To assess the robustness of our findings across different scientific disciplines, we conducted discipline-specific comparisons of innovativeness between PWs and their NPWs. Using the matched sample established in Section 3.1.1, we performed Welch's t-tests within each discipline to compare the two groups across the three innovativeness measures: novelty, convergence, and interdisciplinarity.

These comparisons were performed at three temporal intervals: 20 years before the prize, 5 years before the prize, and after the prize. Table S5 reports the t-statistics for each innovation metric across six representative disciplines, including Chemistry, Physics, Medicine, Biology, Economics, and Geology.

The results show that in most disciplines and across all three measures, PWs demonstrate significantly higher innovativeness than their matched NPWs, particularly in the post-award period. Importantly, even 5-year before receiving the prize, PWs tend to show a modest but consistent advantage, which becomes more pronounced after the award. This pattern reinforces our central findings.

**Table S5. Innovativeness of Prizewinners and Matched Non-prizewinners by Fields.** The table reports t-statistics from Welch's t-tests comparing prizewinners with matched non-prizewinners across three innovativeness measures: novelty, convergence, and interdisciplinarity. Results are stratified by disciplines (e.g., Chemistry, Physics, Biology) and time period relative to award: 20 years before (pre 20y), 5 years before (pre 5y), and post-award (post). Positive values indicate higher scores for PWs.

| | Chemistry | | | Physics | | | Medicine | | |
|---|---|---|---|---|---|---|---|---|---|
| | pre 20y | pre 5y | post | pre 20y | pre 5y | post | pre 20y | pre 5y | post |
| **Novelty** | 4.829*** | 6.358*** | 12.065*** | 6.500*** | 5.765*** | 10.745*** | -0.572 | 0.195 | 4.899*** |
| **Convergence** | 0.648 | 3.845*** | 4.374*** | -0.269 | -2.840** | 1.536 | 1.822 | 2.583** | 9.309*** |
| **Interdisciplinarity** | 1.511 | 4.170*** | 4.327*** | 18.506*** | 14.960*** | 9.074*** | 14.648*** | 9.838*** | 11.874*** |

| | Biology | | | Economics | | | Geology | | |
|---|---|---|---|---|---|---|---|---|---|
| | pre 20y | pre 5y | post | pre 20y | pre 5y | post | pre 20y | pre 5y | post |
| **Novelty** | -2.743** | -1.048 | 3.971*** | 1.552 | 0.435 | 4.835*** | 1.745 | 2.589** | 3.653*** |
| **Convergence** | -0.393 | 3.094** | 5.677*** | 0.560 | 0.229 | 3.869*** | 1.552 | 2.331* | 2.377* |
| **Interdisciplinarity** | 14.967*** | 8.552*** | 11.645*** | 9.236*** | 6.003*** | 7.536*** | 5.787*** | 4.376*** | 3.513*** |

The values in the table are t-statistics from t-tests. Welch's t-test used. Positive t-values indicate higher scores for PWs. *** $p < 0.001$, ** $p < 0.01$, * $p < 0.05$.

To further examine the temporal robustness of our findings, we divided PWs into two cohorts based on the year of their first award: those who received a prize before 1990 and those who were awarded after 1990. For each cohort, we compared the post-award levels of novelty, convergence, and interdisciplinarity between PWs and NPWs using

Welch's t-tests.

As reported in Table S6, the results reveal a consistent pattern across both time periods. PWs exhibit significantly higher levels of novelty and interdisciplinarity relative to their matched counterparts, regardless of whether they received their awards in the earlier or more recent scientific era. The convergence measure also shows positive differences, particularly in the post-1990 group.

**Table S6. Innovativeness Measures Comparison in Different Time Periods.** The table reports t-statistics from Welch's t-tests comparing innovativeness measures between prizewinners and matched non-prizewinners, separately for those who received awards before 1990 and after 1990. Positive t-values indicate higher innovativeness among PWs.

|  | Prizewinning year before 1990 | Prizewinning year after 1990 |
|---|---|---|
|  | t-value | t-value |
| **Novelty** | 8.689*** | 20.490*** |
| **Convergence** | 11.236*** | 13.392*** |
| **Interdisciplinarity** | 4.051*** | 29.429*** |

Welch's t-test used. Positive t-values indicate higher scores for PWs. *** p < 0.001，** p < 0.01，* p < 0.05.

# 4. Regression Models for Innovation Dynamics

## 4.1 Variables Definition

We conducted the fixed-effect ordinary least squares regressions in our analysis. Below are the variables in the models.

**Dependent variables.** We define $\vec{y} = (N, H, R)$, which is a vector of $l*m$ dimension ($l$ is the number of observations, $m = 3$) to represent the three innovation measures.

(1) $N = (n_1, n_2, ..., n_i, ...)$: $n_i$ measures whether a paper $i$ is novel or not.

(2) $H = (h_1, h_2, ... h_i, ...)$: $h_i$ measures whether a paper $i$ is a convergent paper or not.

(3) $R = (r_1, r_2, ... r_i, ...)$: $r_i$ measures the interdisciplinarity of paper $i$.

**Predictors of interest.** A binary variable $w_i$ to indicate whether the author of paper $i$ is a PW or not, $w_i = 1$ if the author of paper $i$ is a PW, $w_i = 0$ otherwise. $p_i$ indicates whether a paper was published after prizewinning or not, $p_i = 1$ if paper $i$ published after prizewinning, $p_i = 0$ otherwise. $k_i$ indicates whether the author of paper $i$ is a multiple prizewinning PW or not, $k_i = 1$ if the PW of paper $i$ is a multiple prizewinning scholar, $k_i = 0$ otherwise. In addition, we consider the interaction term $w_i \times p_i$ to explore the extra benefit of prizewinning in predicting papers' innovation indicators. And we use the interaction term $w_i \times k_i$ to explore the extra benefit of multiple prizewinning in predicting papers' innovation indicators.

**Control variables.** We also include several explanatory variables to control for other possible factors.

- $T_{ti}$: $T_{ti}$ indicates fixed effects that account for the size of a scientific team for paper $i$. We categorize a scientific team into 6 bins: $t = 1, t = 2, t = 3, t = 4, t = 5$ and $t > 5$. $T_{ti} = 1$ if the team size of paper $i$ is in bin $t$ and $T_{ti} = 0$ otherwise.

- $PY_{pyi}$: $PY_{pyi}$ indicates fixed effects for the relative publication year of the paper $i$. We categorize the published years into 9 continuous bins with each for 5 years: $py = -20$, $py = -15$, $py = -10$, $py = -5$, $py = 0, py = 5$, $py = 10$, $py = 15$, $py = 20$. $PY_{pyi} = 1$ if the normalized publication year of paper $i$ is in bin $py$ and $PY_{pyi} = 0$ otherwise.

- $S_{si}$: $S_{si}$ indicates fixed effects for the scientist $s$. $S_{si} = 1$ if a paper $i$ is written by a scientist $s$, $S_{si} = 0$ otherwise.

- $G_{gi}$: $G_{gi}$ indicates fixed effects for the matched group which contains a PW and her/his matched NPW. $G_{gi} = 1$ if a paper $i$ is written by a scientist belonging to group $g$, $G_{gi} = 0$ otherwise.

- $A_{ai}$: $A_{ai}$ indicates fixed effects for the award which contains PW. $A_{ai} = 1$ if the scientist of paper $i$ has received the award $a$, $A_{ai} = 0$ otherwise.

- $AP_{ri}$: $AP_{ri}$ indicates fixed effects for the author position of the focal author in paper $i$. We classify author positions into three categories: first, middle, and last. $AP_{ri} = 1$ if the focal author of paper $i$ holds the author position type $r$ (where $r \in$ {first, middle, last}), and $AP_{ri} = 0$ otherwise.

## 4.2 Testing for Pre-Prizewinning Parallel Trends

We used the DID model to identify the impact of prizewinning on scientific innovation.

Note that a key assumption of the DID is that the PW and NPW groups follow parallel trends before prizewinning [13-15]. We examine the trends in innovation for both groups before prizewinning and investigate whether the two groups are indeed comparable. To do so, we conduct the event study and fit the following equation:

$$y_{mi} = \beta_0 + \beta_l \overline{y}_{-mi} + \sum_{py,py \neq 0} \beta_{pyw} Treat_{pyi} + \sum_t \beta_t T_{ti} + \sum_s \beta_s S_{si} + \sum_{py} \beta_{py} PY_{pyi} + \varepsilon_i, \quad (8)$$

where $Treat_{pyi}$ are a set of dummy variables indicating that if the scientist of paper $i$ prizewinning and paper $i$ was published in year $py$, so that the trend test is not affected by the high volatility of the yearly number of paper innovation. The dummy for $py = 0$ is omitted in Equation (8), so that the post-prizewinning effects are relative to the prizewinning year. From Figure S14, we can see that the corresponding coefficients $\beta_{pyw}$ from -20 years to -5 years before the prizewinning failed to pass the test of significance, indicating that before the prizewinning, there is no significant difference in the variation trend of innovation measures between PWs and matched NPWs groups, which satisfies the parallel trend hypothesis. Therefore, the above estimation of the paper innovation measures is reasonable and feasible.

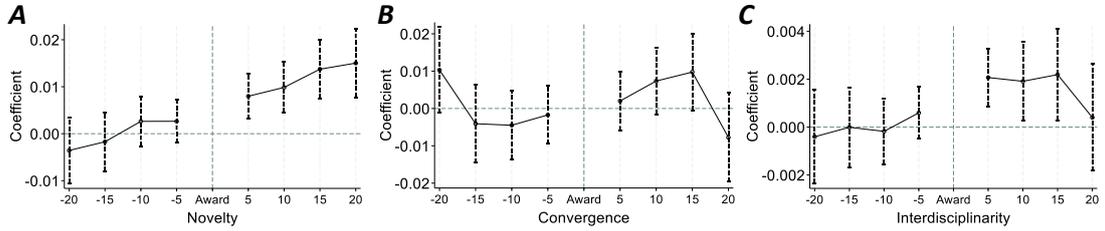

**Figure S14. Testing for Pre-Prizewinning Parallel Trends.** Publication year is normalized relative to the prizewinning. The estimated coefficients and their 95% confidence intervals are plotted. The vertical lines refer to the reference publication year.

## 4.3 Innovation Dynamics and Coevolution

We explored how the innovativeness of the PWs and NPWs coevolve based on a fixed-effect ordinary least squares regression, with controls of prize, team size, matched groups, author position, and publication year. As follows:

$$y_{mi} = \beta_0 + \beta_l \overline{y}_{-mi} + \beta_{wp} w_i \times p_i + \beta_w w_i + \sum_a \beta_a A_{ai} + \sum_t \beta_t T_{ti}$$
$$+ \sum_g \beta_g G_{gi} + \sum_r \beta_r AP_{ri} + \sum_{py} \beta_{py} PY_{pyi} + \varepsilon_i, \quad (9)$$

where $y_{mi}$ represents the $m^{th}$ innovation indicator of $i^{th}$ paper, $\beta_0$ states the regression constant, $\beta_l$ $(l = 1, 2, 3)$ reflects the degree of influence of the other three indicators on the $m^{th}$ indicator for the paper $i$, $\varepsilon_i$ is the regression estimate error.

The full regression results are presented in Figure 3(A) in the main text. The results in models 1-3 demonstrate that there is a strong connection between prizewinning and innovativeness after controlling other explanatory variables, such as team size, matched group, and the other three innovation indicators.

## 4.4 Innovation Dynamics with Time

We explored how the innovation between the PW and NPW groups evolved with time, with other control variables. As follows:

$$y_{mi} = \beta_0 + \beta_l \overrightarrow{y}_{-mi} + \beta_w w_i + \sum_{py} \beta_{py} PY_{pyi} \times w_i + \sum_a \beta_a A_{ai}$$
$$+ \sum_t \beta_t T_{ti} + \sum_g \beta_g G_{gi} + \sum_r \beta_r AP_{ri} + \varepsilon_i. \tag{10}$$

The regression results are shown in Table S7, and the marginal plots based on the regression models are shown in Figure 3(B) in the main text.

**Table S7. Innovation Dynamics with Time.** The independent variables in the three models are the three measures of innovations respectively. "Y" marks the inclusion of corresponding fixed effects in the model.

|  | Model 1 | Model 2 | Model 3 |
|---|---|---|---|
|  | **Novelty** | **Convergence** | **Interdisciplinarity** |
| **Prizewinner (yes=1)** | 0.007** (0.002) | 0.011** (0.004) | 0.006*** (0.001) |
| **Prizewinner × py = -20** | -0.010** (0.003) | 0.004 (0.006) | -0.003** (0.001) |
| **Prizewinner × py = -15** | -0.008** (0.003) | -0.007 (0.005) | -0.002** (0.001) |
| **Prizewinner × py = -10** | -0.001 (0.002) | -0.006 (0.004) | -0.001 (0.001) |
| **Prizewinner × py = 0** | -- | -- | -- |
| **Prizewinner × py = 5** | 0.005 (0.003) | 0.007 (0.005) | 0.003*** (0.001) |
| **Prizewinner × py = 10** | 0.006** (0.002) | 0.015*** (0.005) | 0.001 (0.001) |
| **Prizewinner × py = 15** | 0.010** (0.004) | 0.019*** (0.005) | <-0.001 (0.001) |
| **Prizewinner × py = 20** | 0.011** (0.004) | 0.002 (0.006) | -0.004* (0.002) |
| **Novelty** | -- | -0.001 (0.001) | 0.014*** (0.000) |
| **Convergence** | <-0.001 (0.000) | -- | 0.002*** (0.000) |
| **Interdisciplinarity** | 0.217*** (0.005) | 0.069*** (0.008) | -- |
| **Prize** | Y | Y | Y |
| **Team size** | Y | Y | Y |
| **Matched group** | Y | Y | Y |
| **Author position** | Y | Y | Y |
| **Constant** | 0.044*** (0.001) | 0.425*** (0.002) | 0.190*** (0.000) |
| **Observations** | 2,838,611 | 2,838,611 | 2,838,611 |
| **R-squared** | 0.037 | 0.064 | 0.140 |

## 4.5 Robustness Checks

### 4.5.1 Regressions for Validating the Two Matching Distance Metrics

We used the regression with fixed effects on the two DOM matchings with different distance measures ($d_{i,j}$ and $md_{i,j}$) to validate the conclusion that PWs are associated with publishing more innovative works than their NPWs. The model specification is:

$$y_{mi} = \beta_0 + \beta_{wp} w_i \times p_i + \beta_w w_i + \sum_a \beta_a A_{ai} + \sum_t \beta_t T_{ti}$$
$$+ \sum_g \beta_g G_{gi} + \sum_r \beta_r AP_{ri} + \sum_{py} \beta_{py} PY_{pyi} + \varepsilon_i, \tag{11}$$

where $y_{mi}$ represents the $m^{th}$ innovation indicator of $i^{th}$ paper, $\beta_0$ states the regression constant, $\varepsilon_i$ is the regression estimate error.

The results are presented in Table S8. The regression results are consistent with the trends plot in Figure 2 in the main text (matching with distance $d_{i,j}$) and Figure S10 (matching with distance $md_{i,j}$). We can see that the results from the two matching methods are the same, which means that our conclusions are robust against different matching methods.

**Table S8. Regression Analysis to Confirm Innovation Dynamics Across Different Matching Distance Metrics.** Distance measures $d_{i,j}$ and $md_{i,j}$ indicate the different matching methods we used. "Y" marks the inclusion of corresponding fixed effects in the model.

| | (1) Novelty | | (2) Convergence | | (3) Interdisciplinarity | |
|---|---|---|---|---|---|---|
| | $d_{i,j}$ | $md_{i,j}$ | $d_{i,j}$ | $md_{i,j}$ | $d_{i,j}$ | $md_{i,j}$ |
| **Prizewinner × post** | 0.012*** | 0.014*** | 0.012*** | 0.010** | 0.002 | 0.001 |
| | (0.002) | (0.002) | (0.003) | (0.003) | (0.001) | (0.001) |
| **Prizewinner** | 0.004* | 0.002 | 0.009** | 0.011*** | 0.004*** | 0.005*** |
| | (0.002) | (0.002) | (0.003) | (0.003) | (0.001) | (0.001) |
| **prize** | Y | Y | Y | Y | Y | Y |
| **Team size** | Y | Y | Y | Y | Y | Y |
| **Matched group** | Y | Y | Y | Y | Y | Y |
| **Author position** | Y | Y | Y | Y | Y | Y |
| **Publication year** | Y | Y | Y | Y | Y | Y |
| **Constant** | 0.095*** | 0.094*** | 0.392*** | 0.388*** | 0.186*** | 0.185*** |

| | (0.001) | (0.001) | (0.001) | (0.001) | (0.000) | (0.000) |
|---|---|---|---|---|---|---|
| **Observations** | 3,002,607 | 3,022,185 | 2,993,321 | 3,013,066 | 2,892,826 | 2,908,713 |
| **R-squared** | 0.034 | 0.033 | 0.066 | 0.063 | 0.136 | 0.12 |

Robust standard errors in parentheses, \*\*\*, p<0.001, \*\*, p<0.01, \* p<0.05.

### 4.5.2 Innovation Dynamics with Multiple Prizewinning

We explored how the innovation between the PW and NPW groups evolved with multiple prizewinning of PWs, with other control variables. The fixed-effect ordinary least squares regression with controls of prize, team size, matched groups, author position, and publication year as follows:

$$y_{mi} = \beta_0 + \beta_I \overline{y}_{-mi} + \beta_{wk} w_i \times k_i + \beta_{wp} w_i \times p_i + \sum_a \beta_a A_{ai}$$

$$+ \sum_t \beta_t T_{ti} + \sum_g \beta_g G_{gi} + \sum_r \beta_r AP_{ri} + \sum_{py} \beta_{py} PY_{pyi} + \varepsilon_i, \qquad (12)$$

where $y_{mi}$ represents the $m^{th}$ innovation indicator of $i^{th}$ paper, $\beta_0$ states the regression constant, $\beta_I$ $(I = 1, 2, 3)$ reflects the degree of influence of the other three indicators on the $m^{th}$ indicator for the paper $i$, $\varepsilon_i$ is the regression estimate error.

The full regression results are presented in Table S9. The results in models 1-3 demonstrate that there is an amplified connection between multiple prizewinning and innovativeness after controlling for other explanatory variables, such as team size, matched group, and the other three innovation indicators.

**Table S9. Regression Analysis to Confirm Innovation Dynamics Based on Multiple Prizewinning.** The independent variables in the three models are the three measures of innovativeness respectively. Multi-prizewinning amplified innovative output. "Y" marks the inclusion of corresponding fixed effects in the model.

| | Model 1 | Model 2 | Model 3 |
|---|---|---|---|
| | **Novelty** | **Convergence** | **Interdisciplinarity** |
| **Prizewinner × multi-prizewinning** | 0.005\*\* (0.002) | 0.010\*\* (0.003) | 0.005\*\*\* (0.001) |
| **Prizewinner × post** | 0.011\*\*\* (0.002) | 0.012\*\*\* (0.003) | 0.001 (0.001) |
| **Novelty** | -- | <-0.001 (0.001) | 0.014\*\*\* (0.000) |
| **Convergence** | <-0.001 (0.000) | -- | 0.002\*\*\* (0.000) |
| **Interdisciplinarity** | 0.217\*\*\* (0.005) | 0.067\*\*\* (0.008) | -- |
| **Prize** | Y | Y | Y |
| **Team size** | Y | Y | Y |
| **Matched group** | Y | Y | Y |
| **Author position** | Y | Y | Y |
| **Publication year** | Y | Y | Y |
| **Constant** | 0.055*** (0.001) | 0.386*** (0.002) | 0.184*** (0.000) |
| **Observations** | 2,838,611 | 2,838,611 | 2,838,611 |
| **R-squared** | 0.037 | 0.065 | 0.140 |

Robust standard errors in parentheses, *** p<0.001, ** p<0.01, * p<0.05.

### 4.5.3 Innovation Dynamics after Excluding Review Papers

We examined the innovation dynamics of PWs and NPWs after excluding review papers, using a fixed-effects ordinary least squares regression with controls for prize status, team size, matched groups, author position, and publication year. As follows:

$$y_{mi} = \beta_0 + \beta_I \overline{y}_{-mi} + \beta_{wp} w_i \times p_i + \beta_w w_i + \sum_a \beta_a A_{ai} + \sum_t \beta_t T_{ti}$$
$$+ \sum_g \beta_g G_{gi} + \sum_r \beta_r AP_{ri} + \sum_{py} \beta_{py} PY_{pyi} + \varepsilon_i, \tag{13}$$

where $y_{mi}$ represents the $m^{th}$ innovation indicator of $i^{th}$ paper, $\beta_0$ states the regression constant, $\beta_I$ $(I = 1, 2, 3)$ reflects the degree of influence of the other three indicators on the $m^{th}$ indicator for the paper $i$, $\varepsilon_i$ is the regression estimate error.

The full regression results are presented in Table S10. Models 1–3 show a consistent link between prizewinning and innovativeness, controlling for team size, matched groups, and the other three innovation indicators. The consistency of results after excluding review papers further supports the robustness of our findings.

**Table S10. Innovation Dynamics and Coevolution Excluding Review Papers.** The independent variables in the three models are the three measures of innovativeness respectively. Multi-prizewinning amplified innovative output. "Y" marks the inclusion of corresponding fixed effects in the model.

| | Model 1 | Model 2 | Model 3 |
|---|---|---|---|
| | **Novelty** | **Convergence** | **Interdisciplinarity** |
| **Prizewinner** | 0.003 (0.002) | 0.009** (0.003) | 0.004*** (0.001) |
| **Prizewinner × post** | 0.012*** (0.002) | 0.012*** (0.003) | 0.001 (0.001) |
| **Novelty** | -- | <0.001 (0.001) | 0.014*** (0.000) |
| **Convergence** | <0.001 (0.000) | -- | 0.002*** (0.000) |
| **Interdisciplinarity** | 0.217*** (0.005) | 0.064*** (0.008) | -- |
| **Prize** | Y | Y | Y |
| **Team size** | Y | Y | Y |
| **Matched group** | Y | Y | Y |
| **Author position** | Y | Y | Y |
| **Publication year** | Y | Y | Y |
| **Constant** | 0.055*** (0.001) | 0.386*** (0.002) | 0.184*** (0.000) |
| **Observations** | 2,823,686 | 2,823,686 | 2,823,686 |
| **R-squared** | 0.037 | 0.065 | 0.140 |

### 4.5.4 Staggered Difference-in-Difference Model

We employed the staggered Difference-in-Differences (DID) model as a robustness check to ensure the validity of our findings [14]. It allows for treatment switching in addition to time-varying, heterogeneous treatment effects. By incorporating this method, we address potential biases and enhance the reliability of our results. The consistency of our findings across different estimators, including the staggered DID model, underscores the robustness of our conclusions. The matched prizewinning group in our dataset has inconsistent prizewinning years, presenting multiple queues $c$, then the staggered DID model as follows:

$$y_{mi} = \beta_0 + \beta_1 \overline{y}_{-m,i} + \sum_c \sum_{py} \beta_{cpy} Treat_{c,pyi} + \sum_t \beta_t T_{ti} + \sum_s \beta_s S_{si} + \sum_{py} \beta_{py} PY_{pyi} + \varepsilon_i, \quad (14)$$

where $Treat_{c,pyi}$ are a set of dummy variables indicating that if the scientist of paper $i$ belongs to a queue $c$ and paper $i$ was published in year $py$. $\beta_{c,py}$ represents the average treatment effect (ATE) of queue $c$.

The estimated results in Table S11 indicate that the ATE of novelty and interdisciplinarity measures are statistically significant. This is confirmatory with the findings in our empirical data, as shown in Figure 1 in the text.

**Table S11. Average Cumulative (total) Effect Per Treatment Unit.** The independent variables in the three models are the three measures of innovations respectively.

| | Model 1 | Model 2 | Model 3 |
|---|---|---|---|
| | **Novelty** | **Convergence** | **Interdisciplinarity** |
| **Estimate** | 0.009*** (0.002) | 0.001 (0.004) | 0.002* (0.001) |

### 4.5.5 Random Forest Regression Model

To complement the parametric DID framework and assess the robustness of our findings, we implemented a Random Forest (RF) regression model as a non-linear, non-parametric tool. RF enables us to examine whether post-award innovativeness can be effectively predicted by prizewinning, thereby providing a robustness check under a more flexible machine learning framework.

We trained three separate RF regression models, each targeting a distinct outcome variable—novelty, convergence, and interdisciplinarity. The models were implemented using the RandomForestRegressor class from the scikit-learn package (version 1.5.2) in Python. Each model included the treatment indicator (prizewinner), its interaction with a post-award period indicator (prizewinner × post), the other two innovativeness measures, and a set of one-hot encoded categorical controls, including prize, discipline, team size, relative publication year, and author ID. To ensure computational efficiency while retaining sufficient statistical power, we randomly sampled 1,000 matched groups from the full set constructed in Section 3.1.1. The resulting pooled sample was split into training (70%) and testing (30%) sets. Each RF model was trained using 100 trees, this specification allows for a flexible, non-parametric evaluation of the extent to which post-award innovativeness can be predicted from prizewinning and related covariates. Each RF model was trained using 100 trees, providing a flexible, non-parametric approach that reduces variance and improves stability by averaging over diverse decision trees. This ensemble method yields predictions less sensitive to noise than those from a single tree, with final outputs computed as the mean of all trees' predictions.

To assess the agreement between predicted and observed values, we report the Concordance Correlation Coefficient (CCC) for each model. The CCC simultaneously captures both precision—defined as the correlation between predicted and observed values—and accuracy, which reflects the deviation of the predictions from the line of perfect concordance (i.e., the 45-degree identity line)[16]. CCC is defined as:

$$CCC = \frac{2\rho\sigma_x\sigma_y}{\sigma_x^2 + \sigma_y^2 + (\mu_x - \mu_y)^2},$$ (15)

where $\rho$ is the Pearson correlation coefficient, $\mu_x$, $\mu_y$ are the means, and $\sigma_x^2, \sigma_y^2$ re the variances of predicted and observed values, respectively. Unlike Pearson's $r$ or $R^2$, CCC penalizes both bias and variance, and thus is more appropriate when the goal is to assess overall prediction fidelity, rather than just linear association. A CCC value of 1 indicates perfect agreement, while values closer to 0 indicate poor concordance.

In our context, CCC allows us to directly assess whether the predicted post-award innovativeness measures from different models align with observed outcomes not only in direction, but also in magnitude and distribution. The consistently high CCC scores observed across both DID and RF models confirm that the observed treatment effects are not artifacts of model specification and reinforce the reliability of our main conclusion.

Figure S15 compares the predicted values of the three post-award innovativeness measures from the DID and RF models against the observed outcomes. Panels A–C display predictions from the DID model versus actual values. The CCC indicates moderately strong agreement, particularly for convergence (CCC = 0.70) and interdisciplinary (CCC = 0.86). Panels D–F show the RF model predictions, which yield comparable or slightly higher CCC values, especially for novelty and interdisciplinary. Panels G–I compare the predictions of the DID and RF models directly, revealing high consistency between the two approaches (CCC = 0.88, 0.85, and 0.94, respectively).

These results confirm that the observed treatment effects are not artifacts of the linear assumptions of DID, and remain robust under a flexible, ensemble-based learning framework.

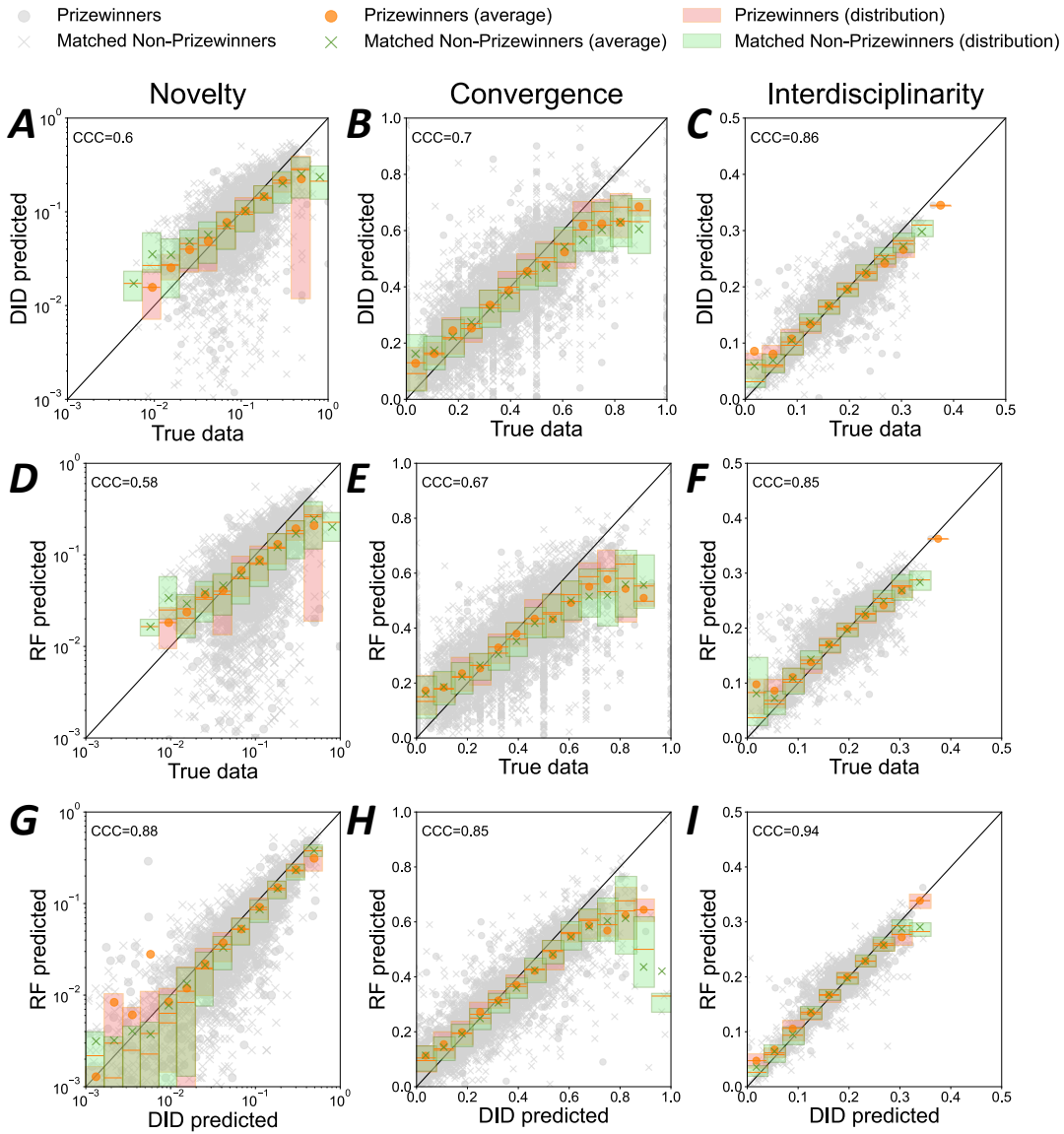

**Figure S15. Comparison of Prediction Agreement for Random Forest, DID Regression, and Actual data across Three Innovativeness Measures.** (A–C) show scatterplots comparing the predicted values from the difference-in-differences (DID) model with true observed values for novelty, convergence, and interdisciplinarity, respectively. (D–F) show the same comparison for the RF model, and (G–I) directly compare predictions from the DID and RF models. Each point represents a matched group of prizewinners (orange) or matched non-prizewinners (gray), with corresponding average values overlaid as orange circles (prizewinners) and green crosses (matched non-prizewinners). Colored bands indicate the distribution of predicted values for each group. Concordance Correlation Coefficient (CCC) is shown in each plot.

# 5. Embeddedness and Innovation

## 5.1 Network Embeddedness Definition

Here we examined how PWs differ from NPWs in maintaining their collaboration networks and sustaining their research topic trajectories over the course of their careers.

We focused on the dynamic structure of collaboration and topic evolution, employing three key metrics: tie duration, tie overlap, and topic similarity. The first two metrics capture relational behaviors in collaboration maintenance, while the third reflects knowledge embeddedness in research topics. Figure S16 illustrates the schematic workflow of our analytical approach.

- Tie duration represents how long the scholar keeps each of her/his collaboration ties. The collaboration tie duration is the average collaboration time length between the scholar and each of the collaborators within a focal paper (see Figure S16(A)).

- Tie overlap measures the degree of interaction between collaborators' collaborative networks within a focal paper. Each scholar's network tie overlap measured by the average Jaccard similarity values (see Figure S16(B)).

- Topic similarity quantifies the differences in research diversity between PWs and NPWs. This measure assesses the similarity between the scholar's pre-research topics and the topic of the focal paper. Here we used the fine-scale concept classification from OpenAlex (OpenAlex assigns each paper with hierarchical concept tags using machine learning technique), which enables us to capture topic differences that would be obscured at a coarser scale (see Figure S16(C)).

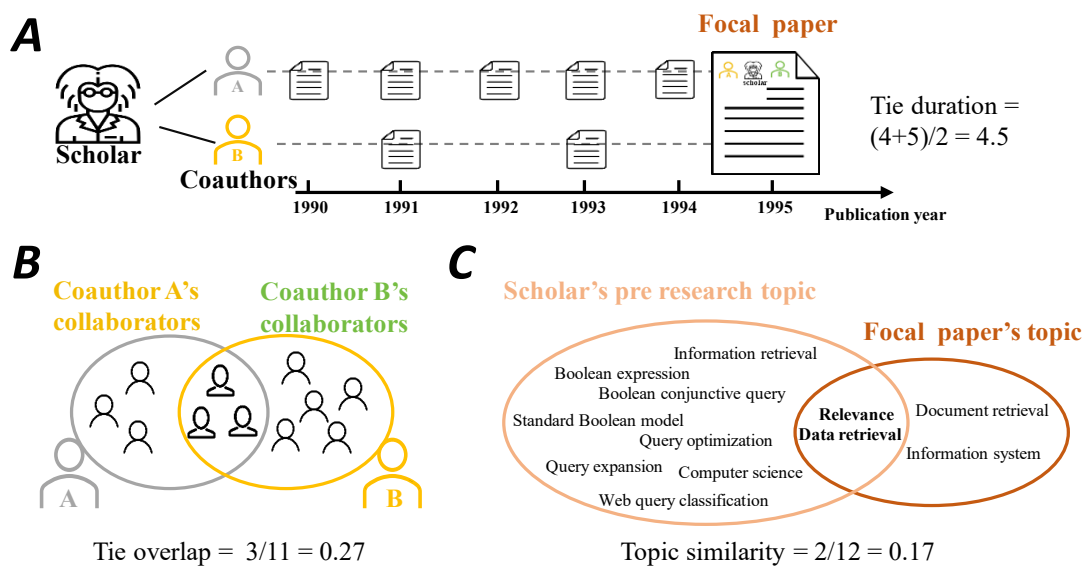

**Figure S16. Network Embeddedness.** (A) Iillustration of a scholar's collaboration timeline prior to the publication of a paper in 1995. Tie duration is defined as the average number of years a scholar collaborates with each coauthor on a given paper. (B) Tie overlap is the average Jaccard similarity over all pairs of coauthors within the focal paper, where the Jaccard similarity assesses the overlapping rates between coauthors' coauthors. (C) Topics are defined according to the concepts in OpenAlex. The topic similarity between the focal collaborating paper and the scholar's prior research foci is similarly determined using the Jaccard similarity index.

## 5.2 Prizewinner's Network Embeddedness and Innovation

### 5.2.1 Variables Definition

We conducted the fixed-effect ordinary least squares regressions in this part. Below are the variables in the models.

**Dependent variables.** We define $\vec{z} = (\boldsymbol{z_d}, \boldsymbol{z_o}, \boldsymbol{z_s})$, which is a vector of $k*m$ dimension ($m = 3$), used to represent the tie duration, tie overlap, and topic similarity.

(1) $\boldsymbol{z_d} = (z_{d1}, z_{d2}, \ldots z_{di}, \ldots)$: $z_{di}$ measures the collaboration network tie duration of paper $i$.

(2) $\boldsymbol{z_o} = (z_{o1}, z_{o2}, \ldots z_{oi}, \ldots)$: $z_{oi}$ measures the tie overlap of paper $i$.

(3) $\boldsymbol{z_s} = (z_{s1}, z_{s2}, \ldots z_{si}, \ldots)$: $z_{si}$ measures the topic similarity of paper $i$.

**Predictors of interest.** We use a binary variable $w_i$ to indicate whether the scholar is a PW or not, $w_i = 1$ if a scholar $i$ is a PW, $w_i = 0$ otherwise. $p_i$ indicates whether a paper was published after prizewinning or not, $p_i = 1$ if paper $i$ published after prizewinning, $p_i = 0$ otherwise. In addition, we consider the interaction term $w_i \times p_i$ to explore the extra benefit of prizewinning in predicting papers' innovation indicators.

**Control variables.**

- $T_{ti}$: $T_{ti}$ indicates fixed effects that account for the size of a scientific team for paper $i$. We categorize a scientific team into 6 bins: $t = 1, t = 2, t = 3, t = 4, t = 5$ and $t > 5$. $T_{ti} = 1$ if the team size of paper $i$ is in bin $t$ and $T_{ti} = 0$ otherwise.

- $PY_{pyi}$: $PY_{pyi}$ indicates fixed effects for the relative publication year of the paper $i$. We categorize the published years into 9 continuous bins with each for 5 years: $py = -20,\ py = -15,\ py = -10,\ py = -5,\ py = 0, py = 5,\ py = 10,\ py = 15,\ py = 20$. $PY_{pyi} = 1$ if the normalized publication year of paper $i$ is in bin $py$ and $PY_{pyi} = 0$ otherwise.

- $G_{gi}$: $G_{gi}$ indicates fixed effects for the matched group which contains a PW and her/his matched NPW. $G_{gi} = 1$ if a paper $i$ is written by a scientist belonging to group $g$, $G_{gi} = 0$ otherwise.

- $A_{ai}$: $A_{ai}$ indicates fixed effects for the award which contains PW. $A_{ai} = 1$ if the scientist of paper $i$ has received the award $a$, $A_{ai} = 0$ otherwise.

- $AP_{ri}$: $AP_{ri}$ indicates fixed effects for the author position of the focal author in paper $i$. We classify author positions into three categories: first, middle, and last. $AP_{ri} = 1$ if the focal author of paper $i$ holds the author position type $r$ (where $r \in$ {first, middle, last}), and $AP_{ri} = 0$ otherwise.

### 5.2.2 Model Specifications

**(1) Regression models for network embeddedness variables.**

To complement the descriptive patterns in Figure 4, we estimated fixed-effects regression models to assess whether PWs exhibit significantly different levels of network embeddedness compared to NPWs. Specifically, we modeled three dimensions of embeddedness—tie duration, tie overlap, and topic similarity—at the paper level using the following specification:

$$z_{mi} = \beta_0 + \beta_w w_i + \beta_{wp} w_i \times p_i + \beta_p p_i + \sum_a \beta_a A_{ai}$$
$$+ \sum_t \beta_t T_{ti} + \sum_g \beta_g G_{gi} + \varepsilon_i. \qquad (16)$$

where $z_{mi} = (z_{di}, z_{oi}, z_{si})$ represents the tie duration, tie overlap, and topic similarity of a scholar's $i^{th}$ paper, respectively.

The regression results are presented in Table S12. In all three models, the coefficients on the *Prizewinner* indicator are negative and highly significant (p<0.001), indicating that prizewinners exhibit consistently lower levels of embeddedness than matched NPWs across all dimensions. Specifically, prizewinners are associated with shorter tie duration, lower tie overlap, and lower topic similarity. These findings are consistent with the descriptive trends observed in Figure 4.

Furthermore, the interaction term *Prizewinner×post* is significantly negative for tie duration, suggesting that the gap in tie duration between PWs and NPWs further widens following prize receipt. However, the interaction effects for tie overlap and topic similarity are close to zero and not statistically significant, implying stability in those dimensions after award.

**Table S12. Innovativeness and the network embeddedness.** Each model controls for the fixed effects of prize, team size, PWs and NPWs matched groups, prize, and team size.

|  | Model 1 | Model 2 | Model 3 |
|---|---|---|---|
|  | Tie duration | Tie overlap | Topic similarity |
| **Prizewinner** | -0.185*** (0.033) | -0.007*** (0.001) | -0.003*** (0.000) |
| **Prizewinner × post** | -0.170*** (0.041) | <-0.001 (0.001) | <0.001 (0.000) |
| **Post** | 1.963*** (0.017) | -0.023*** (0.001) | -0.029*** (0.000) |
| **Prize** | Y | Y | Y |
| **Matched group** | Y | Y | Y |
| **Team size** | Y | Y | Y |
| **Constant** | 3.732*** (0.016) | 0.099*** (0.001) | 0.053*** (0.000) |
| **Observations** | 4,622,989 | 3,232,356 | 5,485,438 |
| **R-squared** | 0.108 | 0.098 | 0.242 |

Robust standard errors in parentheses, *** p<0.001, ** p<0.01, * p<0.05.

## (2) Relationship between network embeddedness and innovativeness.

Our analysis indicates that differences in network embeddedness predict innovativeness. To determine the relationship between collaborative characteristics and innovation between PWs and NPWs, we used the fixed effects regression models, as follows:

$$y_{mi} = \beta_0 + \beta_I z_{mi} + \beta_w w_i + \beta_{wp} w_i \times p_i + \sum_a \beta_a A_{ai}$$
$$+ \sum_t \beta_t T_{ti} + \sum_g \beta_g G_{gi} + \sum_r \beta_r AP_{ri} + \sum_{py} \beta_{py} PY_{pyi} + \varepsilon_i, \qquad (17)$$

where $y_{mi}$ represents the $m^{th}$ innovation measure of $i^{th}$ paper (defined in SI Sec.4.1), and $z_{mi} = (z_{di}, z_{oi}, z_{si})$ represents the tie duration, tie overlap, and topic similarity of paper $i$, respectively. $\beta_0$ states the regression constant, $\beta_I$ $(I = 1, 2, 3)$ reflects the degree of influence of the network embeddedness variables on the $m^{th}$ indicator for the paper $i$, $\varepsilon_i$ is the regression estimate error.

The corresponding regression outcomes are shown in Table 1 in the main text. The regression results confirmed the strong predictive relationship between network embeddedness and innovativeness. Collectively, PWs are embedded in collaborations that are relatively short, involve coauthors whose third-party ties have low overlap and work on research topics that significantly differ from the PW's prior topics.

## 5.3 The productivity and citation impact of the coauthors of PWs and NPWs

To examine whether differences in relative innovativeness between PWs and NPWs could be attributed to disparities in the productivity or impact of their coauthors, we constructed two focal-paper-level measures reflecting coauthors' prior scientific performance.

Specifically, for each focal paper authored by a PW or NPW, we identified all coauthors excluding the focal individual. For each coauthor, we retrieved their cumulative number of publications and total citations from OpenAlex up to, but not including, the year of publication of the focal paper. These values were then averaged across all coauthors of the focal paper, yielding two variables:

- Coauthors' #papers: the average number of publications produced by coauthors prior to the focal paper;

- Coauthors' #citations: the average number of citations accumulated by coauthors prior to the focal paper.

These variables serve as proxies for the prior productivity and citation impact of the collaborative environment surrounding each focal paper. As shown in Figure S17, the distributions of these two indicators do not differ significantly between the coauthors of PWs and NPWs, supporting the conclusion that the observed differences in innovativeness are not driven by systematic differences in the research caliber of their collaborators.

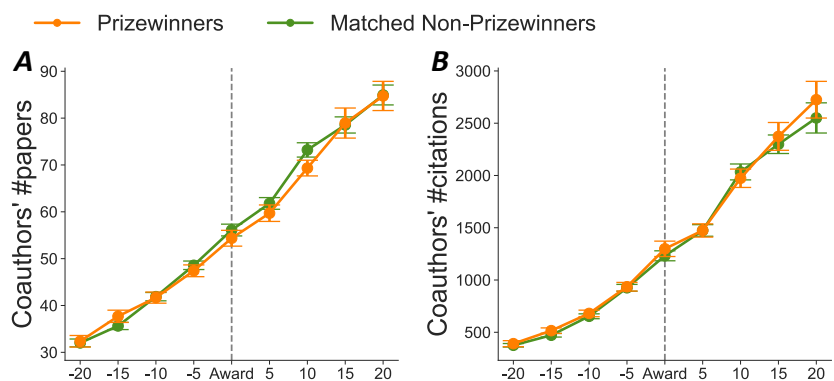

**Figure S17. Comparison of coauthors' productivity and citation impact for PWs and NPWs.** Average number of publications (A) and citations (B) by coauthors of PWs and NPWs from 20 years before to 20 years after the prizewinning year. Dots denote the averages, and error bars represent the 95% confidence intervals. No significant

differences are observed between the two groups, suggesting that the observed differences in innovativeness are not attributable to systematic differences in the productivity or citation impact of their coauthors.